\documentclass[preprint,12pt]{elsarticle}
\usepackage{graphicx}
\usepackage{amssymb}
\usepackage{amsmath}
\usepackage{subfig,color}
\usepackage{multirow}
\usepackage{colortbl,color}
\usepackage{algorithm}
\usepackage{multirow}
\usepackage{rotating}
\usepackage{amsthm}
\usepackage{hyperref}
\biboptions{comma,sort&compress}
\journal{ }

\begin{document}

\begin{frontmatter}
\title{ Upscaled Lattice Boltzmann Method for Simulations of Flows in Heterogeneous Porous Media}
\author[kaustNP]{Jun~Li}
\author[kaustNP]{Donald~Brown}
%
\address[kaustNP]{Center for Numerical Porous Media\\King Abdullah University of Science and Technology\\Thuwal, Saudi Arabia}
\begin{abstract}

A upscaled lattice Boltzmann method (LBM) for flow simulations in heterogeneous porous media, at both pore and Darcy scales, is proposed in this paper. In the micro-scale simulations, we model flows using LBM with the modified Guo et al. algorithm where we replace the force model with a simple Shan-Chen force model.
The proposed upscaled LBM uses coarser grids to represent the effects of the fine-grid (pore-scale) simulations. For the upscaled LBM, effective properties and reduced-order models are proposed  as we coarsen the grid. The effective properties are computed using solutions of local problems (e.g., by performing local LBM simulations) subject to some boundary conditions. A upscaled LBM that can reduce the computational complexity of existing LBM and transfer the information between different scales is implemented. The results of coarse-grid, reduced-order, simulations agree very well with averaged results obtained using a fine grid.

\end{abstract}
\end{frontmatter}
\section{Introduction}\label{s:intro}

Detailed flow simulations in porous media are often modeled using the Darcy or Brinkman approximations. In these models, effective parameters, such as absolute and relative permeabilities, depend on the pore-scale geometry. To compute these effective parameters, pore-scale simulations accounting for relevant geometric features in a Representative Elementary Volume (REV) are commonly used as in \cite{Khan2012:porescale app}.  The lattice Boltzmann method (LBM) \cite{McNamara1988:LBM}-\cite{Qian1992:D2Q9} is well developed for pore-scale flow simulations and extended to model two-phase systems or two immiscible fluids \cite{Shan1993:S-C}-\cite{Shan1996:S-C review}.  After computing the effective parameters, we are able to perform Darcy-scale simulations using traditional finite volume or element methods used in commercial reservoir simulators. However, these computations are limited to small REVs (compared to the computational grid) and rely on two distinct idealized scale concepts.

Flows at the Darcy scale can also be modeled by LBM with a modified algorithm. The model described in \cite{Zhu2013:Darcy} allows particles to partially bounce back at the cells (points) with small permeability. In \cite{Kang2002:Darcy}-\cite{Guo2002:Darcy}, an external body force, which increases with decreasing permeability, is employed to represent the resistance effect of the porous media to the fluid, where LBM is considered as a unified framework for simulations at all scales. However, these simulations require significant computational resources to converge since the permeability distribution usually has drastic changes in space, which requires a very fine grid for high spatial resolution. To overcome this difficulty, we propose a upscaled LBM scheme that is applicable at the pore and coarser, e.g., Darcy scales.

Following the work in \cite{Guo2002:Darcy} where the generalized Navier-Stokes equation \cite{Nithiarasu1997:GeneralizedN-S} is solved, we simplify the equilibrium distribution function. In addition, we replace the original Guo et al. force model \cite{Guo2002:forcemodel} used in \cite{Guo2002:Darcy} by the simpler Shan-Chen force model \cite{Shan1993:S-C}-\cite{Shan1994:S-C} to improve the efficiency.
Then, an upscaled LBM scheme is proposed to improve computational efficiency by using a coarse grid (each coarse point represents a subdomain) with effective permeability. For each subdomain, the effective permeability is computed by a local scheme, which is based on the conservation principle for the average fluxes (see \cite{Efendiev2009:MsFEM} for general overview of multiscale methods). To avoid the iterative process of finding the unknown effective permeability that satisfies the equation for the average flux, we derive an analytical formula. This analytical formula allows finding the average flux in terms of the effective permeability, which is then inversely determined from the computed average flux using the original permeability distribution in the subdomain concerned.

The computed effective permeabilities are verified in several benchmark problems, where analytical solutions are known. We implement upscaled LBM simulations using a coarse grid and the computed effective permeability. Agreement between the coarse and fine grid LBM simulations demonstrates the validity of the upscaled LBM scheme. The average effects of the fine-grid simulations are preserved in the coarse-grid simulations in solving the flow equation at any intermediate scale. Our numerical results show that one can achieve a substantial gain in CPU time by using coarse-grid models. In this paper, the upscaled LBM approach is applied to single-phase flows; however, this approach can be used for modeling multi-phase flow phenomena.
\section{LBM algorithms for simulating flows in porous media}\label{s:LBM algorithm for Brinkman}

In this section, we discuss LBM algorithms that will be used in our microscale simulations.
We first present LBM algorithm based on the force model proposed by Guo et al. in \cite{Guo2002:forcemodel}, where an additional term is used in the particle evolution equation to represent the force contribution. Then, we present the general Shan-Chen force model for multicomponent and multiphase systems. In our upscaling algorithm, we focus on the single phase and single component model for Brinkmann flows. The Shan-Chen force model allows for a more efficient upscaling procedure and overall cleaner presentation.
We refer to \cite{Sukop2006:LBMbook} for more general discussions on LBM algorithms.

First, we will introduce some basic notation associated with LBM. The grid (lattice) points are uniformly distributed inside the computational domain and $c=\Delta x/\Delta t$, where $\Delta x$ is the lattice spacing and $\Delta t$ is the time step. For two-dimensional problems, we use the D2Q9 lattice model \cite{Qian1992:D2Q9}, where $Q=9$ is the total number of lattice velocities, $\vec e_0=(0, 0)$, $\omega_0=4/9$, $\vec e_\alpha=(\cos\theta_\alpha, \sin\theta_\alpha)c$, $\theta_\alpha=(\alpha-1)\pi/2$, $\omega_\alpha=1/9$ for $\alpha=$ 1 to 4, $\vec e_\alpha=(\cos\theta_\alpha, \sin\theta_\alpha)\sqrt{2}c$, $\theta_\alpha=(\alpha-5)\pi/2+\pi/4$, and $\omega_\alpha=1/36$ for $\alpha=$ 5 to 8. For three-dimensional problems, the D3Q19 lattice model \cite{Qian1992:D2Q9} with different $\vec e_\alpha$ and $\omega_\alpha$ is used, but the algorithms are unchanged.
\subsection{LBM algorithm with the Guo force model}\label{ss:original LBM algorithm}
Following the algorithm presented in  \cite{Guo2002:Darcy}, the evolution algorithm of the distribution function $f_\alpha(\vec x, t)$ is:

\begin{equation}\label{eq:f evolution of Guo}
\begin{aligned}
    f_\alpha(\vec x+\Delta t\vec e_\alpha, t+\Delta t)=f_\alpha(\vec x, t)+\dfrac{f_\alpha^{\rm (eq)}(\vec x, t)-f_\alpha(\vec x, t)}{\tau}+\Delta tF_\alpha(\vec x, t),
\end{aligned}
\end{equation}
where the normalized relaxation time $\tau$ is appropriately selected to match the desired effective kinematic viscosity $\nu_{\rm eff}=c_{\rm s}^2(\tau-0.5)\Delta t$, where $c_{\rm s}=c/\sqrt{3}$ is the sound speed. We use the simplified truncated form of the  equilibrium distribution function as:

\begin{equation}\label{eq:feq}
\begin{aligned}
    f_\alpha^{\rm (eq)}=\omega_\alpha\rho(1+\dfrac{\vec e_\alpha\cdot\vec u^{\rm (eq)}}{c_{\rm s}^2}),
\end{aligned}
\end{equation}
where the density is given by $\rho(\vec x, t)=\sum_{\alpha=0}^{Q-1}f_\alpha(\vec x, t)$ and the equilibrium velocity, $\vec u^{\rm (eq)}(\vec x, t),$ is defined as:

\begin{equation}\label{eq:ueq of Guo}
\begin{aligned}
    \vec u^{\rm (eq)}=\dfrac{\sum_{\alpha=0}^{Q-1}{\vec e_\alpha f_\alpha}+\dfrac{1}{2}\Delta t\rho\vec f_{\rm m}}{\rho},
\end{aligned}
\end{equation}
where $\vec f_{\rm m}(\vec x, t)$ is the force per unit mass. Similarly, $F_\alpha(\vec x, t)$ is simplified to:

\begin{equation}\label{eq:Fi of Guo}
\begin{aligned}
    F_\alpha=\omega_\alpha\rho(1-\dfrac{1}{2\tau})\dfrac{\vec e_\alpha\cdot\vec f_{\rm m}}{c_{\rm s}^2}.
\end{aligned}
\end{equation}

In the force model proposed by Guo et al. \cite{Guo2002:forcemodel}, the flow velocity $\vec u$ is equal to $\vec u^{\rm (eq)}$. If $\vec f_{\rm m}$ is constant, $\rho$ and $\vec u^{\rm (eq)}$ are computed by $f_\alpha$ and then $f_\alpha^{\rm (eq)}$ and $F_\alpha$ of Eq. \eqref{eq:f evolution of Guo} are determined explicitly. For solving the pore-scale equation, we consider the following expressions for $\vec f_{\rm m}$ as a linear function of $\vec u$ (see \cite{Guo2002:Darcy}):

\begin{equation}\label{eq:f_m depends on u}
\begin{aligned}
    \vec f_{\rm m}=-\dfrac{\epsilon\nu}{\kappa}\vec u+\epsilon\vec G,
\end{aligned}
\end{equation}
where $\epsilon$ is the porosity, $\nu$ is the physical kinematic viscosity of the fluid, $\kappa(\vec x)$ is a scalar for the permeability and $\vec G(\vec x)$ is the external body force per unit mass. The force introduced above incorporates the porous media heterogeneities through the permeability function $\kappa(\vec x)$ and depends on the microstructure. If $\kappa(\vec x)$ has a high value in the region, then one can assume that this region is highly permeable, while if
$\kappa(\vec x)$  has a very low value, then this region is almost impermeable. One can also use a forcing that is nonlinear in $\vec u$ as an extension to cases with  nonlinear Forchheimer effects which are discussed in \cite{Guo2002:Darcy} and \cite{Li2013:SPE}.

In expressions \eqref{eq:ueq of Guo} and \eqref{eq:f_m depends on u} we have $\vec u=\vec u^{\rm (eq)}$. Using this fact and solving for $\vec u$ in \eqref{eq:f_m depends on u}, we obtain the explicit formula as in  \cite{Guo2002:Darcy}:

\begin{equation}\label{eq:explicit u in Guo}
\begin{aligned}
    \vec u&=\dfrac{\sum_{\alpha=0}^{Q-1}\vec e_\alpha f_\alpha+\dfrac{\Delta t}{2}\epsilon\rho\vec G}{\rho(1+\dfrac{\epsilon\Delta t\nu}{2\kappa})}.
\end{aligned}
\end{equation}
$\vec f_m$ is computed by Eq. \eqref{eq:f_m depends on u}. Then, $f_\alpha^{\rm (eq)}$ and $F_\alpha$ of Eq. \eqref{eq:f evolution of Guo} are determined by  \eqref{eq:feq} and \eqref{eq:Fi of Guo}, respectively.

In the incompressible limit with $|\vec u|\ll c_{\rm s}$, the analysis \cite{Guo2002:Darcy} based on the Chapman-Enskog expansion shows that the computed pressure $p=c_{\rm s}^2\rho$ and flow velocity $\vec u$ converge to the solutions of the following equation:

\begin{equation}\label{eq:Brinkman-like}
\begin{aligned}
    \nabla\cdot\vec u&=0 \\
    \dfrac{\partial\vec u}{\partial t}&=-\dfrac{1}{\rho_0}\nabla p+\nu_{\rm eff}\Delta\vec u-\dfrac{\epsilon\nu}{\kappa}\vec u+\epsilon\vec G,
\end{aligned}
\end{equation}
where $\rho_0$ is the initial mass density used in LBM simulations. Here, $\rho_0$ needs not be the real density $\rho_{\rm real}$ of the incompressible fluid; then, the computed $p\rho_{\rm real}/\rho_0$ is used as the pressure of the physical problem. The steady state results of LBM simulations are used as the solutions of the Brinkman equation. The parameters of $\nu_{\rm eff}$, $\epsilon$, $\nu$ and $\kappa(\vec x)$ can be set independently such that the steady state LBM results converge to the solutions of the continuum Darcy and Stokes equations, respectively.
\subsection{Simplified LBM algorithm with the Shan-Chen force model}\label{ss:simple LBM algorithm}
We now present the general Shan-Chen model and its application to our upscaling scheme. In the original Shan-Chen model \cite{Shan1993:S-C}-\cite{Shan1994:S-C}, which is proposed to simulate multiphase and multicomponent flows, the number of molecules of the $\sigma^{\rm th}$ component having the velocity $\vec e_\alpha$ at $\vec x$ and time $t$ is denoted by $f_\alpha^\sigma(\vec x, t)$, where $\sigma=1, \cdots , S$ and $S$ is the total number of components. The general updating algorithm of $f_\alpha^\sigma(\vec x, t)$ is:

\begin{equation}\label{eq: general f evolution of S-C}
\begin{aligned}
    f_\alpha^\sigma(\vec x+\Delta t\vec e_\alpha, t+\Delta t)=f_\alpha^\sigma(\vec x, t)+\dfrac{f_\alpha^{\sigma({\rm eq})}(\vec x, t)-f_\alpha^\sigma(\vec x, t)}{\tau^\sigma},
\end{aligned}
\end{equation}
where $\sigma=1, \cdots, S$ and the equilibrium distribution function is defied as:

\begin{equation}\label{eq: general feq of S-C}
\begin{aligned}
    f_\alpha^{\sigma({\rm eq})}=\rho^\sigma\omega_\alpha\Bigg(1+\dfrac{\vec e_\alpha\cdot\vec u^{\sigma({\rm eq})}}{c_{\rm s}^2}+\dfrac{(\vec e_\alpha\cdot\vec u^{\sigma({\rm eq})})^2}{2c_{\rm s}^4}-\dfrac{\vec u^{\sigma({\rm eq})}\cdot\vec u^{\sigma({\rm eq})}}{2c_{\rm s}^2}\Bigg),
\end{aligned}
\end{equation}
where $\sigma=1, \cdots, S$, $\rho^\sigma=\sum_{\alpha=0}^{Q-1} f_\alpha^\sigma$ and $\vec u^{\sigma({\rm eq})}$ is computed as:

\begin{equation}\label{eq: general ueq of S-C}
\begin{aligned}
    \vec u^{\sigma({\rm eq})}=\dfrac{\rho^\sigma\vec u^\prime +\tau^\sigma\vec F^\sigma}{\rho^\sigma},   \qquad\sigma=1, \cdots , S,
\end{aligned}
\end{equation}
where $\vec F^\sigma(\vec x, t)$ is related to the total volume force acting on the $\sigma^{\rm th}$ component. Generally speaking \cite{Kang2002:F123}, $\vec F^\sigma$ contains three parts: the fluid-fluid interaction $\vec F^{1,\sigma}$, fluid-solid interaction $\vec F^{2,\sigma}$ and external force $\vec F^{3,\sigma}$. For example, $\vec F^{3,\sigma}=\Delta t\rho^\sigma\vec G$ for the contribution by the external body force $\vec G$ per unit mass. In Eq. \eqref{eq: general ueq of S-C}, $\vec u^\prime$ is defined as follows to conserve momentum:

\begin{equation}\label{eq: general uprime of S-C}
\begin{aligned}
    \vec u^\prime=\dfrac{\sum_{\sigma=1}^S\dfrac{1}{\tau^\sigma}\sum_{\alpha=0}^{Q-1}\vec e_\alpha f_\alpha^\sigma}{\sum_{\sigma=1}^S\dfrac{1}{\tau^\sigma}\sum_{\alpha=0}^{Q-1}f_\alpha^\sigma}.
\end{aligned}
\end{equation}
The flow velocity $\vec u$ of the whole fluid is equal to the mean velocity before and after implementing the force term and is computed as follows:

\begin{equation}\label{eq: general uflow of S-C}
\begin{aligned}
    \vec u=\dfrac{\sum_{\sigma=1}^S\sum_{\alpha=0}^{Q-1}\vec e_\alpha f_\alpha^\sigma+\dfrac{1}{2}\sum_{\sigma=1}^S\vec F^\sigma}{\sum_{\sigma=1}^S\rho^\sigma}.
\end{aligned}
\end{equation}
Recently \cite{Li2013:SPE}, phase separation process in a fiber geometry and flow of two immiscible fluids in a cross channel are modeled using the Shan-Chen model, which shows the convenience of the LBM in dealing with complex geometries and manipulating the contact angle.

As upscaling in the multiphase phase is a very difficult and often nonlinear procedure, we focus our algorithm first to single-component and single-phase models. For flows of single-component and single-phase, the evolution of $f_\alpha(\vec x, t)$ without notation $\sigma$ is:
\begin{equation}\label{eq:f evolution of S-C}
\begin{aligned}
    f_\alpha(\vec x+\Delta t\vec e_\alpha, t+\Delta t)=f_\alpha(\vec x, t)+\dfrac{f_\alpha^{\rm (eq)}(\vec x, t)-f_\alpha(\vec x, t)}{\tau}.
\end{aligned}
\end{equation}
In order to recover the Brinkman equation, the equilibrium distribution function $f_\alpha^{\rm (eq)}$ of Eq. \eqref{eq: general feq of S-C} is simplified to the above Eq. \eqref{eq:feq}. $\rho=\sum_{\alpha=0}^{Q-1}f_\alpha$ but $\vec u^{\rm (eq)}$ of Eq. \eqref{eq: general ueq of S-C} is modified as follows:

\begin{equation}\label{eq:ueq of S-C}
\begin{aligned}
    \vec u^{\rm (eq)}=\dfrac{\sum_{\alpha=0}^{Q-1}{\vec e_\alpha f_\alpha}+\tau\Delta t\rho\vec f_{\rm m}}{\rho},
\end{aligned}
\end{equation}
where we use $\Delta t\rho\vec f_{\rm m}$ to replace the original notation which is equal to the momentum increase per unit volume after $\Delta t$ due to the force effect through the relaxation process. Correspondingly, the flow velocity $\vec u$ of Eq. \eqref{eq: general uflow of S-C} is modified to:

\begin{equation}\label{eq:u of S-C}
\begin{aligned}
    \vec u=\dfrac{\sum_{\alpha=0}^{Q-1}{\vec e_\alpha f_\alpha}+\dfrac{1}{2}\Delta t\rho\vec f_{\rm m}}{\rho}.
\end{aligned}
\end{equation}
When solving the Brinkman equation, $\vec f_{\rm m}$ is a function of $\vec u$ defined by Eq. \eqref{eq:f_m depends on u}. A comparison of Eqs. \eqref{eq:ueq of Guo} and \eqref{eq:u of S-C} shows that the explicit formula of Eq. \eqref{eq:explicit u in Guo} to compute $\vec u$ is also valid here. Then, $\vec u^{\rm (eq)}$ is computed as:
\begin{equation}\label{eq:final ueq of S-C}
\begin{aligned}
    \vec u^{\rm (eq)}=2\tau\vec u+(1-2\tau)\dfrac{\sum_{\alpha=0}^{Q-1}{\vec e_\alpha f_\alpha}}{\rho},
\end{aligned}
\end{equation}
which is obtained by solving Eqs. \eqref{eq:ueq of S-C} and \eqref{eq:u of S-C}.

As we can see, the computations of $\vec f_{\rm m}$ by Eq. \eqref{eq:f_m depends on u} and $F_\alpha$ by Eq. \eqref{eq:Fi of Guo} using the computed $\vec f_{\rm m}$ in the original algorithm \cite{Guo2002:Darcy} are avoided in the current simplified algorithm and, therefore, the efficiency is improved. In the incompressible limit with $|\vec u|\ll c_{\rm s}$, the computed pressure $p=c_{\rm s}^2\rho$ and flow velocity $\vec u$ also converge to the solutions of the above Brinkman-like equation Eq. \eqref{eq:Brinkman-like}. The same idea can be implemented to Eqs. \eqref{eq: general f evolution of S-C}-\eqref{eq: general uflow of S-C} to solve multiphase flows at the Darcy scale. In this way, it may be possible to develop multiphase upscaling techniques based on the upscaling scheme presented below. This is a topic for future work.

\section{Upscaling scheme}\label{s:upscaling scheme}

For many practical cases, the number of fine discretization points in the whole computational domain due to heterogeneities is very large, making the memory usage and computational time unaffordable. We use an upscaling simulation scheme to reduce the number of points in the fine grid by using a coarse grid with an effective permeability $\mbox{\boldmath$\kappa$}^*(\vec x)$. The upscaled quantities are a tensor quantity even though the input permeability, $\kappa(\vec x)$, is assumed to be a scalar. With this approach we are able to capture fine grid information on the coarse grid by solving many parallel local problems.

\begin{figure}
  \centering
  \includegraphics[width=0.49\textwidth]{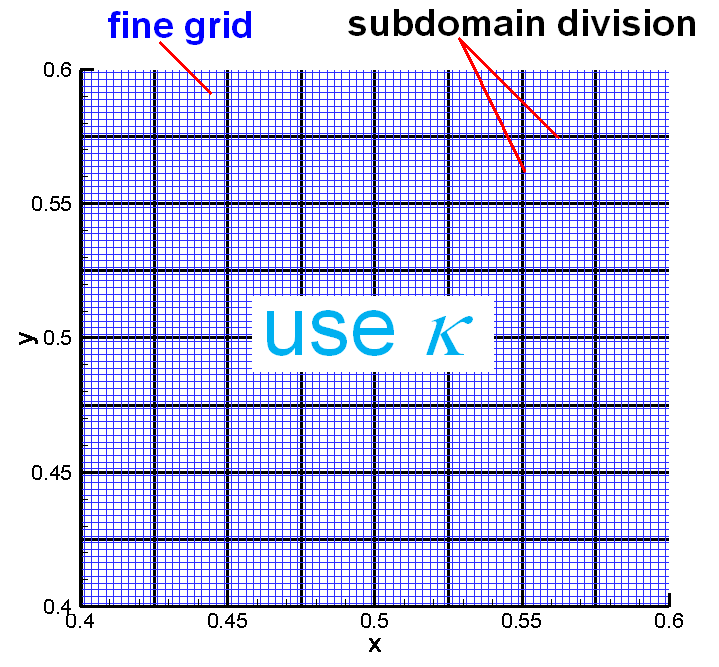}
  \includegraphics[width=0.49\textwidth]{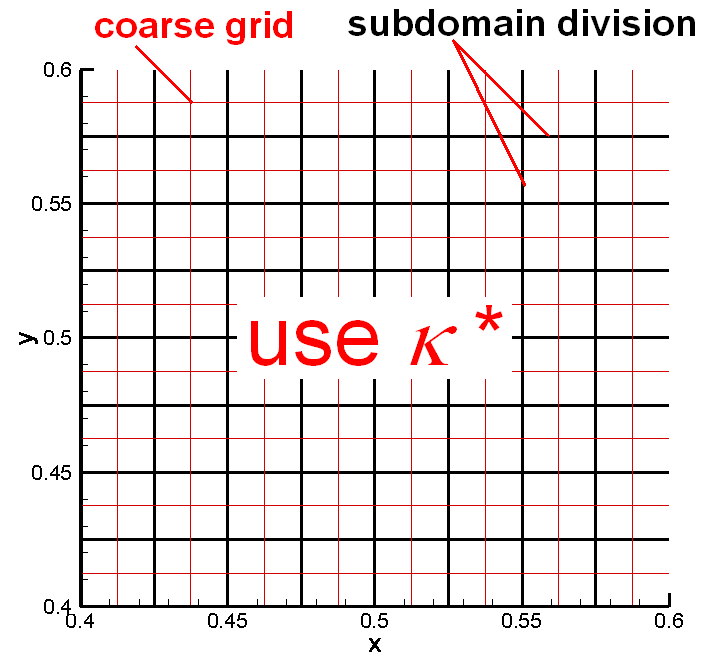}
  \caption{Schematic models of the fine and coarse grids.}
  \label{fig:two grid model}
\end{figure}

In our proposed algorithm, the computational domain is divided into many subdomains and each subdomain is represented by a coarse point (see Fig. \ref{fig:two grid model}). This substantially reduces the degrees of freedom in the coarse-grid simulation. To compute the effective $\mbox{\boldmath$\kappa$}^*$ for each subdomain, we impose different external forces $\vec G_{\rm const}$ to drive flows in different directions in the local LBM simulations, which use a fine grid located inside the corresponding subdomain and the distribution of $\kappa(\vec x)$ on the fine grid. Then, the similar local LBM simulations usually need to be run with a constant tensor $\mbox{\boldmath$\kappa$}^*$ as shown in Eq. \eqref{eq:f_m depends on u and tensor k}. We seek $\mbox{\boldmath$\kappa$}^*$ such that the average velocities from local fine-grid simulations with the heterogeneous $\kappa(\vec x)$ and homogeneous $\mbox{\boldmath$\kappa$}^*$, respectively, are equal (see Eqs. \eqref{eq:conserve u(1)}-\eqref{eq:conserve u(2)}). The onerous seeking process by adjusting the unknown $\mbox{\boldmath$\kappa$}^*$ to match the fluxes computed using $\kappa(\vec x)$ is avoided in our simulations since $\mbox{\boldmath$\kappa$}^*$ can be computed explicitly by Eq. \eqref{eq:tensor kappa}.

We discuss two-dimensional problems as example. In the local LBM simulations using $\kappa(\vec x)$, we drive flow in the $x$ direction by $\vec G^{(1)}_{\rm const}=(G_{\rm const}, 0)$ and compute the average velocity $\overline{\vec u}^{(1)}_{\kappa(\vec x)}$, where $\overline{\cdot}$ is defined as a volume average over a subdomain. We also compute $\overline{\vec u}^{(2)}_{\kappa(\vec x)}$ by using $\vec G^{(2)}_{\rm const}=(0, G_{\rm const})$ in another local simulation. Then, $\mbox{\boldmath$\kappa$}^*$ is computed as follows:

\begin{equation}\label{eq:tensor kappa}
\begin{aligned}
    \mbox{\boldmath$\kappa$}^*={\kappa^*_{xx}, \qquad  \kappa^*_{xy} \choose \kappa^*_{yx}, \qquad  \kappa^*_{yy}}=\dfrac{\nu}{G_{\rm const}}{\overline{\vec u}^{(1)}_{\kappa(\vec x)}\cdot(1, 0), \qquad \overline{\vec u}^{(2)}_{\kappa(\vec x)}\cdot(1, 0) \choose \overline{\vec u}^{(1)}_{\kappa(\vec x)}\cdot(0, 1), \qquad \overline{\vec u}^{(2)}_{\kappa(\vec x)}\cdot(0, 1)}
\end{aligned}
\end{equation}

Now, we validate that the computed $\mbox{\boldmath$\kappa$}^*$ satisfies the conservation principle of average fluxes. Assuming that we run local LBM simulations using the constant $\mbox{\boldmath$\kappa$}^*$ computed by Eq. \eqref{eq:tensor kappa}, Eq. \eqref{eq:f_m depends on u} is modified to be:

\begin{equation}\label{eq:f_m depends on u and tensor k}
\begin{aligned}
    \vec f_{\rm m}=-\epsilon\nu{\mbox{\boldmath$\kappa$}^*}^{-1}\cdot\vec u+\epsilon\vec G,
\end{aligned}
\end{equation}
where ${\mbox{\boldmath$\kappa$}^*}^{-1}$ is the inverse matrix of $\mbox{\boldmath$\kappa$}^*$. The evolution of $f_\alpha(\vec x, t)$ is described by Eqs. \eqref{eq:feq}, \eqref{eq:f evolution of S-C},  \eqref{eq:ueq of S-C}, \eqref{eq:u of S-C} and \eqref{eq:f_m depends on u and tensor k}. As $\mbox{\boldmath$\kappa$}^*$ and $\vec G$ are constant and the periodic boundary conditions are used in local simulations, the relation $$f_\alpha(\vec x+\Delta t\vec e_\alpha, t+\Delta t)=f_\alpha(\vec x, t)$$ holds at steady state. For arbitrary $\Delta x$, $\Delta t$, $\tau$, $\epsilon$, $\nu$, $\mbox{\boldmath$\kappa$}^*$ and $\vec G=\vec G_{\rm const}$, the steady state solution of $f_\alpha$ is independent of $\vec x$ and equal to:

\begin{equation}\label{eq:f steady}
\begin{aligned}
    f_\alpha=\omega_\alpha\rho_0(1+\dfrac{\vec e_\alpha}{c_{\rm s}^2}\cdot\dfrac{\mbox{\boldmath$\kappa$}^*\cdot\vec G_{\rm const}}{\nu})
\end{aligned}
\end{equation}
which implies that the uniform density is $\rho=\sum_{\alpha=0}^{Q-1}f_\alpha\equiv\rho_0$. We validate the solution of Eq. \eqref{eq:f steady} by the following verification: substituting Eq. \eqref{eq:f steady} into Eq. \eqref{eq:u of S-C} and considering Eq. \eqref{eq:f_m depends on u and tensor k}, we get the uniform velocity $\vec u\equiv\dfrac{\mbox{\boldmath$\kappa$}^*\cdot\vec G_{\rm const}}{\nu}$. In addition, we get $\vec f_{\rm m}\equiv0$ by Eq. \eqref{eq:f_m depends on u and tensor k} using $\vec u$. Then, substituting $f_\alpha$ and $\vec f_{\rm m}$ into Eq. \eqref{eq:ueq of S-C}, we get $\vec u^{\rm (eq)}=\vec u$, which implies that Eq. \eqref{eq:f evolution of S-C} is satisfied at steady state since we have $f_\alpha=f_\alpha^{\rm (eq)}$ according to Eq. \eqref{eq:feq} and $f_\alpha(\vec x+\Delta t\vec e_\alpha, t+\Delta t)=f_\alpha(\vec x, t)$. According to the uniform solution of $\vec u\equiv\dfrac{\mbox{\boldmath$\kappa$}^*\cdot\vec G_{\rm const}}{\nu}$ and Eq. \eqref{eq:tensor kappa}, the average velocity $\overline{\vec u}^{(1)}_{\mbox{\boldmath$\kappa$}^*}$ of the local simulation using constant $\mbox{\boldmath$\kappa$}^*$ and $\vec G^{(1)}_{\rm const}=(G_{\rm const}, 0)$ satisfies:

\begin{equation}\label{eq:conserve u(1)}
\begin{aligned}
    \overline{\vec u}^{(1)}_{\mbox{\boldmath$\kappa$}^*}=\dfrac{\mbox{\boldmath$\kappa$}^*\cdot(G_{\rm const}, 0)}{\nu}=\overline{\vec u}^{(1)}_{\kappa(\vec x)}
\end{aligned}
\end{equation}
which implies that the average flux is conserved when using the same external force $\vec G^{(1)}_{\rm const}$ but different permeability distributions, namely using the heterogeneous $\kappa(\vec x)$ and homogeneous $\mbox{\boldmath$\kappa$}^*$, respectively. When driving flow by $\vec G^{(2)}_{\rm const}=(0, G_{\rm const})$, the average flux is also conserved:

\begin{equation}\label{eq:conserve u(2)}
\begin{aligned}
    \overline{\vec u}^{(2)}_{\mbox{\boldmath$\kappa$}^*}=\dfrac{\mbox{\boldmath$\kappa$}^*\cdot(0, G_{\rm const})}{\nu}=\overline{\vec u}^{(2)}_{\kappa(\vec x)}
\end{aligned}
\end{equation}

After getting the value of $\mbox{\boldmath$\kappa$}^*(\vec x)$ at each coarse point on the coarse grid, we implement two LBM simulations on the coarse and fine grids, respectively, inside the whole computational domain. The $\mbox{\boldmath$\kappa$}^*(\vec x)$, $\Delta x_{\rm coarse}$ and $\Delta t_{\rm coarse}$ in the coarse-grid simulation are different from $\kappa(\vec x)$, $\Delta x_{\rm fine}$ and $\Delta t_{\rm fine}$, respectively. The boundary conditions and the parameters $\rho_0$, $\nu_{\rm eff}$, $\epsilon$, $\nu$ and $\vec G(\vec x)$ in the coarse-grid simulation are the same as in the fine-grid simulation. In order to clearly verify the validity of the coarse-grid simulation of the whole computational domain, we use periodic boundary conditions to eliminate potential numerical errors which occur when using fixed pressures, for example, at the two ends along the $x$ direction because fixed quantities are numerically imposed at the initial and last points along the $x$ direction and their spatial positions are different when using different $\Delta x$.

\section{Numerical results}\label{s:numerical results}
\subsection{Comparison between the original and the proposed LBM algorithms}\label{ss:Compare two LBM algorithms}
\begin{figure}
  \centering
  \includegraphics[width=0.32\textwidth]{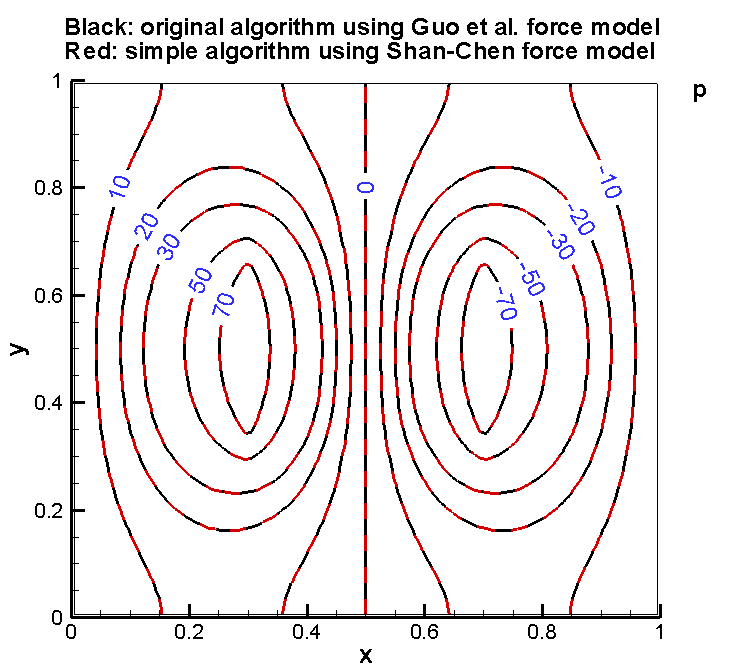}
  \includegraphics[width=0.32\textwidth]{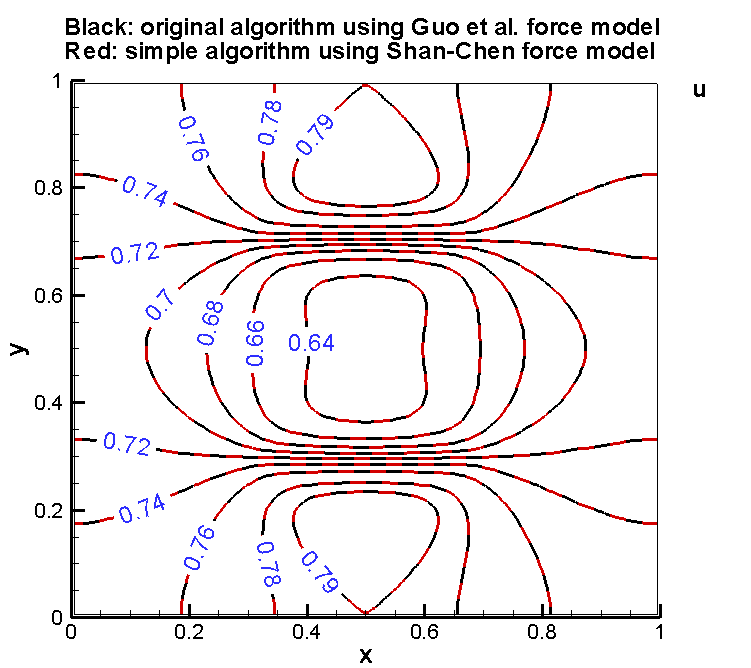}
  \includegraphics[width=0.32\textwidth]{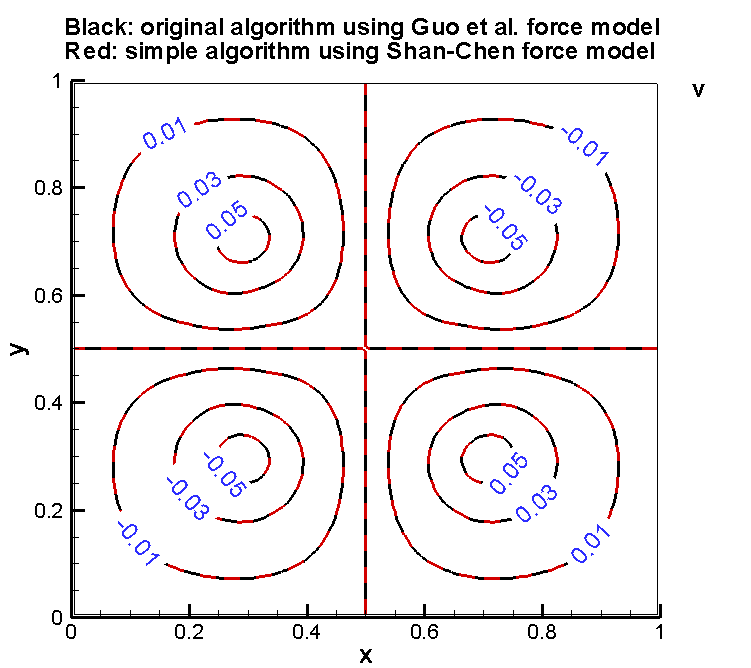}\\
  \includegraphics[width=0.32\textwidth]{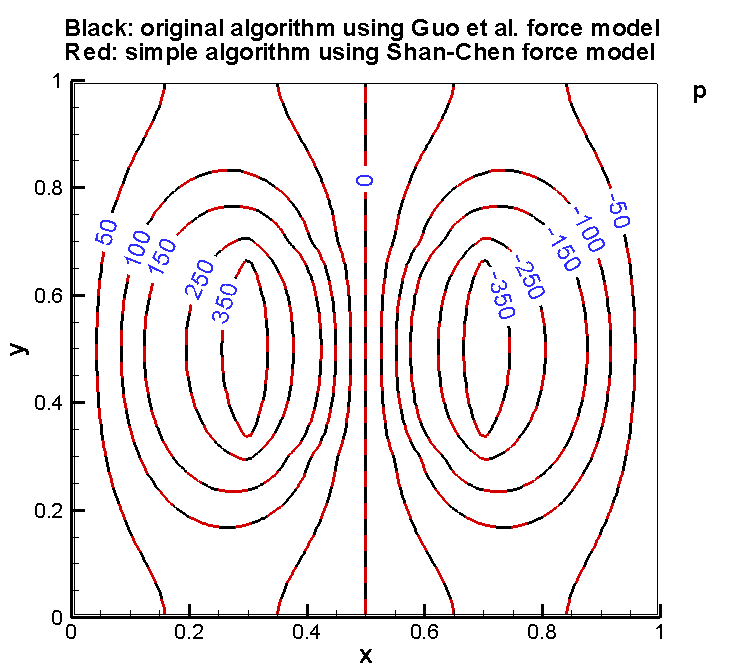}
  \includegraphics[width=0.32\textwidth]{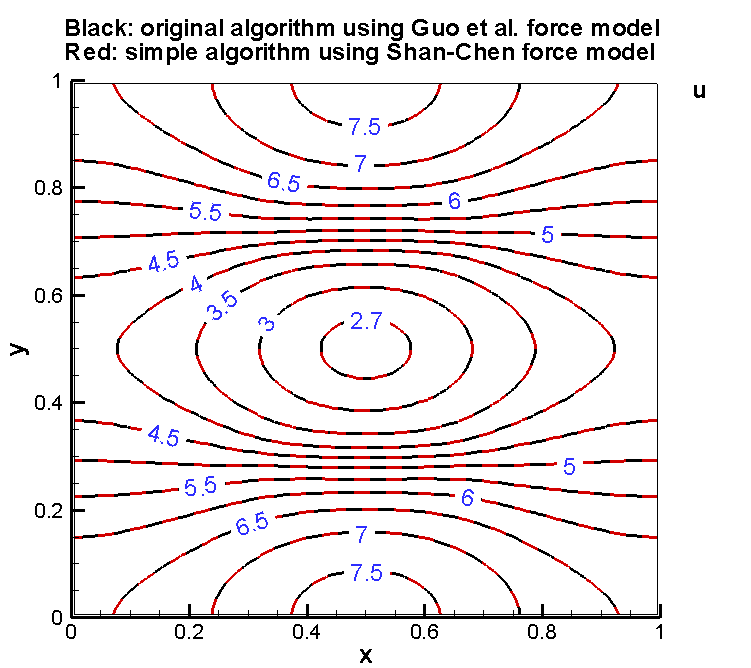}
  \includegraphics[width=0.32\textwidth]{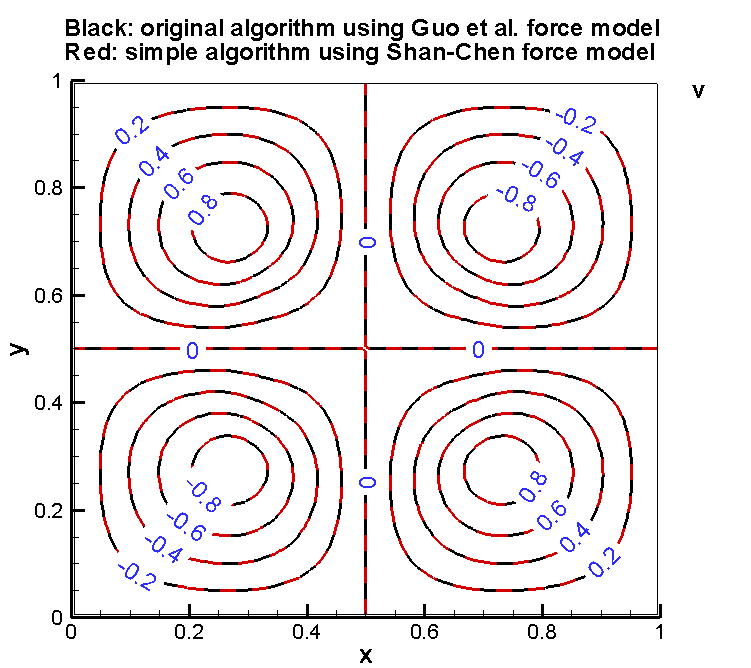}
  \caption{Comparisons of $p$, $u$ and $v$ between two LBM simulations using different force models, transient results at the $5000^{\rm th} \Delta t$ (top) and the steady state results (bottom), $\nu_{\rm eff}=0.01$ m$^2$ s$^{-1}$, $\vec G_{\rm const}=(2, 0)$ m s$^{-2}$.}
  \label{fig:check force model}
\end{figure}

First, we verify the proposed simple LBM algorithm using the Shan-Chen force model against the original algorithm using the Guo el al. force model. In the two simulations using different force models, the number of grid points is $100\times100$ and $\Delta x=0.01$ m, $\Delta t=0.0001$ s and $\tau=0.53$ making $\nu_{\rm eff}=0.01$ m$^2$ s$^{-1}$, $\nu=2\times10^{-6}$ m$^2$ s$^{-1}$, $\rho_0=1000$ kg m$^{-3}$, $\epsilon=0.8$ and $\vec G_{\rm const}=(2, 0)$ m s$^{-2}$. The periodic boundary conditions are used and the permeability assigned to each point with index $(i, j)$ is:

\begin{equation}\label{eq:kappa-forcemodel}
    \begin{cases}\kappa=10^{-6}, & 31\le i,j\le70 \\
    \kappa=10^{-5}, & {\rm otherwise}.
    \end{cases}
\end{equation}
The average pressure over the whole computational domain is subtracted from the computed pressure $p=c_{\rm s}^2\rho$ in all figures of the pressure distributions. The transient results at the $5000^{\rm th} \Delta t$ and the steady state results at the $200000^{\rm th} \Delta t$ are given in Fig. \ref{fig:check force model} which shows the excellent agreement between the two simulations using different force models. The simulation using the Guo et al. force model takes about 23 minutes of computational time but the simulation using the Shan-Chen force model uses about 21 minutes. In the following LBM simulations, we only use the simple LBM algorithm with the Shan-Chen force model, which is described in Section \ref{ss:simple LBM algorithm}.

\subsection{Verifications of the computed $\mbox{\boldmath$\kappa$}^*$}\label{ss:verify kappa star}
We use the above setting of parameters but choose different values for $\tau$, $\vec G_{\rm const}$ and $\kappa$ for different problems in Section \ref{ss:verify kappa star}. We run simulations in the whole computational domain with prescribed distribution of $\kappa(\vec x)$ and verify the computed effective permeability $\mbox{\boldmath$\kappa$}^*$ against the analytical solutions.
\subsubsection{Layered distribution of $\kappa$}\label{sss:layered}
Here, we uniformly divide the whole domain into 10 layers parallel to the $y$ axis. The odd number layers have $\kappa_1=10^{-12}$ m$^2$ and
$\kappa_2$ in the even number layers is constant with values shown in Table \ref{layered gx tau0.53} for different cases. Flow is driven in the $x$ direction by a uniform $\vec G_{\rm const}=(2, 0)$ m s$^{-2}$. $\tau=0.53$ and so $\nu_{\rm eff}=0.01$ m$^2$ s$^{-1}$. The results in Table \ref{layered gx tau0.53} show that the computed $\kappa^*_{xx}$ by LBM agrees exactly with the analytical solution although the analytical formula is derived from the Darcy equation. This is because the steady state velocity is uniform and so the LBM simulations based on the Brinkman equation with nonzero $\nu_{\rm eff}$ actually yield the solutions of the Darcy equation at steady state.

\begin{table}
\caption{Verification of computed $\kappa^*_{xx}$, $\kappa_1=10^{-12}$ m$^2$ and $\nu_{\rm eff}=0.01$ m$^2$ s$^{-1}$}\label{layered gx tau0.53}
\begin{center}
\begin{tabular}{ccc}
\hline
$\dfrac{\kappa_2}{\kappa_1}$ & $[\dfrac{1}{2}(\dfrac{1}{\kappa_1}+\dfrac{1}{\kappa_2})]^{-1}$ & $\kappa^*_{xx}$ by LBM \\
\hline
$2     $                     & $1.33333\times10^{-12}$                                        & $1.33333\times10^{-12}$ \\
$10    $                     & $1.81818\times10^{-12}$                                        & $1.81818\times10^{-12}$ \\
$50    $                     & $1.96078\times10^{-12}$                                        & $1.96078\times10^{-12}$ \\
$100   $                     & $1.98019\times10^{-12}$                                        & $1.98019\times10^{-12}$ \\
$1000  $                     & $1.99800\times10^{-12}$                                        & $1.99800\times10^{-12}$ \\
$10000 $                     & $1.99980\times10^{-12}$                                        & $1.99979\times10^{-12}$ \\
$100000$                     & $1.99998\times10^{-12}$                                        & $1.99998\times10^{-12}$ \\
\hline
\end{tabular}
\end{center}
\end{table}

When driving flow in the $y$ direction by setting $\vec G_{\rm const}=(0, 2)$ m s$^{-2}$, the velocity distribution along the $x$ direction is nonuniform. We set $\tau=0.5$ such that $\nu_{\rm eff}=0$ m$^2$ s$^{-1}$ to recover the Darcy equation. As we can see in Table \ref{layered gy tau0.5}, the computed $\kappa^*_{yy}$ by LBM simulations agrees exactly with the analytical solution.

\begin{table}
\caption{Verification of computed $\kappa^*_{yy}$, $\kappa_1=10^{-12}$ m$^2$ and $\nu_{\rm eff}=0$ m$^2$ s$^{-1}$}\label{layered gy tau0.5}
\begin{center}
\begin{tabular}{ccc}
\hline
$\dfrac{\kappa_2}{\kappa_1}$ & $\dfrac{1}{2}(\kappa_1+\kappa_2)$  & $\kappa^*_{yy}$ by LBM  \\
\hline
$2     $                     & $1.500000\times10^{-12}$           & $1.499999\times10^{-12}$ \\
$10    $                     & $5.500000\times10^{-12}$           & $5.499999\times10^{-12}$ \\
$50    $                     & $25.50000\times10^{-12}$           & $25.49999\times10^{-12}$ \\
$100   $                     & $50.50000\times10^{-12}$           & $50.49999\times10^{-12}$ \\
$1000  $                     & $500.5000\times10^{-12}$           & $500.4999\times10^{-12}$ \\
$10000 $                     & $5000.500\times10^{-12}$           & $5000.499\times10^{-12}$ \\
$100000$                     & $50000.50\times10^{-12}$           & $50000.49\times10^{-12}$ \\
\hline
\end{tabular}
\end{center}
\end{table}
\subsubsection{Checkerboard distribution of $\kappa$}\label{sss:checkerboard}
\begin{figure}
  \centering
  \includegraphics[width=0.32\textwidth]{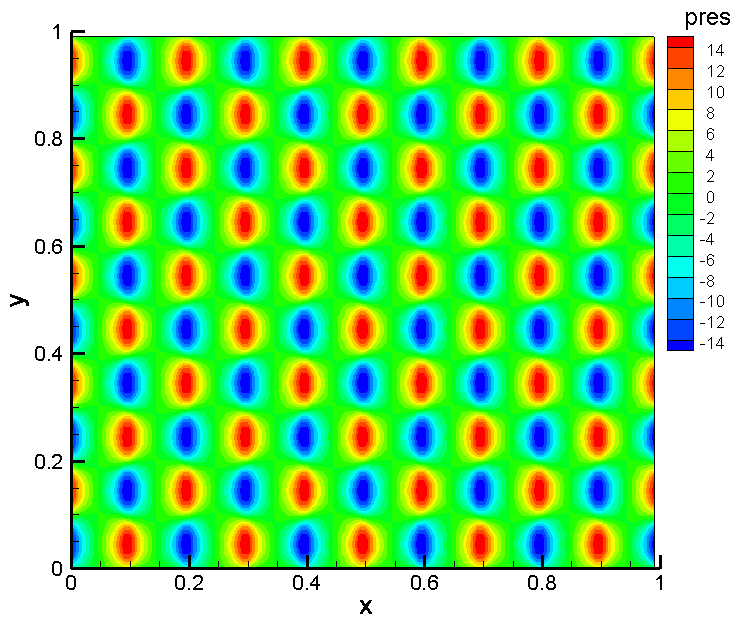}
  \includegraphics[width=0.32\textwidth]{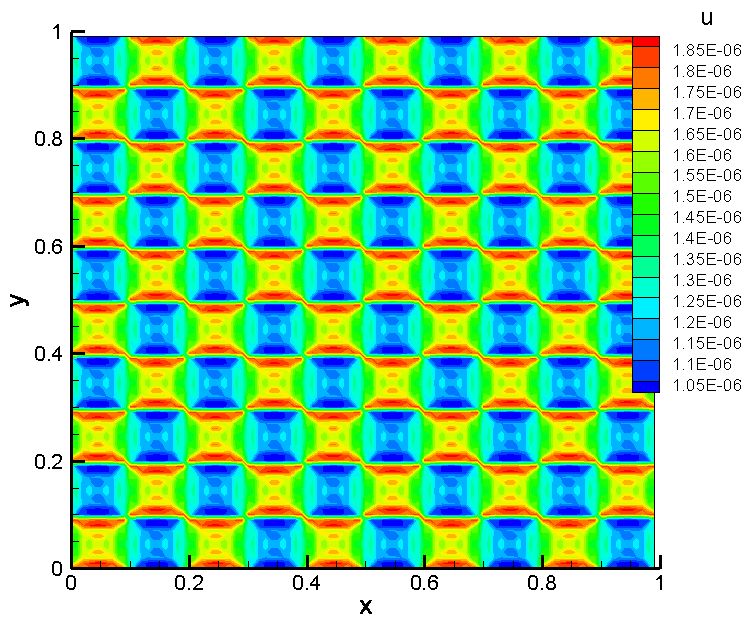}
  \includegraphics[width=0.32\textwidth]{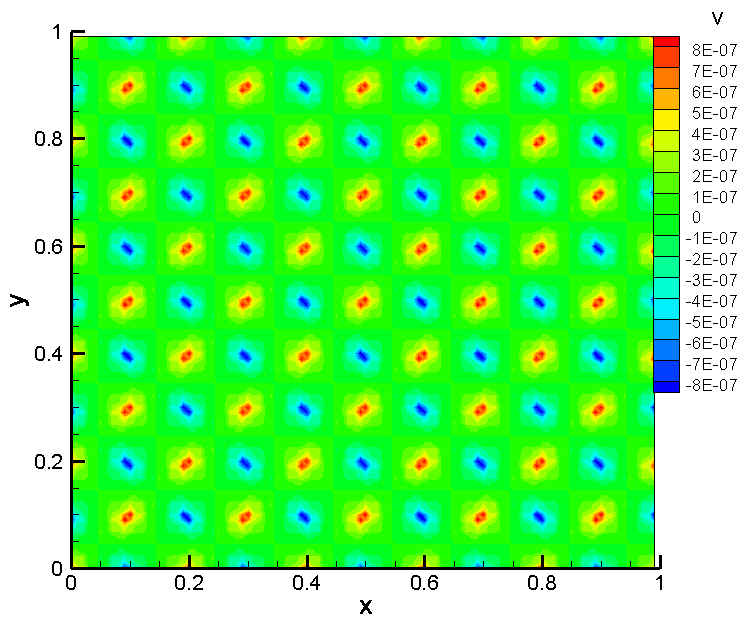}
  \caption{Distributions of $p$, $u$ and $v$, $\nu_{\rm eff}=0$ m$^2$ s$^{-1}$, $\vec G_{\rm const}=(2, 0)$ m s$^{-2}$, $\kappa_1=10^{-12}$ m$^2$ and $\dfrac{\kappa_2}{\kappa_1}=2$.}
  \label{fig:checkerboard uvp}
\end{figure}

As on a checkerboard, we divide the whole computational domain uniformly into $10\times10$ squares with each square containing $10\times10$ points. The black squares of the checkerboard have $\kappa_1=10^{-12}$ m$^2$ and $\kappa_2$ in the white squares takes different values for different cases as shown in Table \ref{checkerboard gx tau0.5}. Flow is driven by a uniform $\vec G_{\rm const}=(2, 0)$ m s$^{-2}$ and we set $\nu_{\rm eff}=0$ m$^2$ s$^{-1}$ to get the solution of the Darcy equation. The representative distributions of $p$, $u$ and $v$ are given in Fig. \ref{fig:checkerboard uvp}. The results in Table \ref{checkerboard gx tau0.5} show that the computed $\kappa^*_{xx}$ by LBM simulations agrees well with the analytical solution when $\dfrac{\kappa_2}{\kappa_1}$ is not very large but deviates significantly in the case of high contrast. This deviation is due to the low spatial resolution of the grid used in the LBM simulations at high contrast of permeability. We refine the grid by increasing the total point number from $100\times100$ to $1000\times1000$ to show improving accuracy. $\Delta x$ and $\Delta t$ are changed to $10^{-3}$ m and $10^{-5}$ s, respectively. The results given in Table \ref{checkerboard gx tau0.5} show that the computed $\kappa^*_{xx}$ becomes very close to the analytical solution when the permeability ratio is up to 100 but still significantly deviate from the correct value if the permeability ratio is very high, where more points are required to achieve good spatial resolution.

\begin{table}
\caption{Verification of computed $\kappa^*_{xx}$, $\kappa_1=10^{-12}$ m$^2$ and $\nu_{\rm eff}=0$ m$^2$ s$^{-1}$}\label{checkerboard gx tau0.5}
\begin{center}
\begin{tabular}{cccc}
\hline
$\dfrac{\kappa_2}{\kappa_1}$ & $\sqrt{\kappa_1\kappa_2}$ & $\kappa^*_{xx}$ by LBM  & $\kappa^*_{xx}$ by LBM ($1000\times1000$ points) \\
\hline
$2     $                     & $1.41421\times10^{-12}$   & $1.41418\times10^{-12}$ &                                          \\
$10    $                     & $3.16227\times10^{-12}$   & $3.14081\times10^{-12}$ &                                          \\
$50    $                     & $7.07106\times10^{-12}$   & $6.45938\times10^{-12}$ & $7.01357\times10^{-12}$                  \\
$100   $                     & $10.0000\times10^{-12}$   & $8.25393\times10^{-12}$ & $9.70489\times10^{-12}$                  \\
$1000  $                     & $31.6227\times10^{-12}$   & $12.2496\times10^{-12}$ & $19.8897\times10^{-12}$                  \\
$10000 $                     & $100.000\times10^{-12}$   & $13.0133\times10^{-12}$ & $23.2777\times10^{-12}$                  \\
\hline
\end{tabular}
\end{center}
\end{table}

\subsection{Verifications of the upscaled simulation scheme}\label{ss:verify upscaling}
\subsubsection{Simulations of Darcy flows}\label{sss:Darcy two grids}
\begin{figure}
  \centering
  \includegraphics[width=0.7\textwidth]{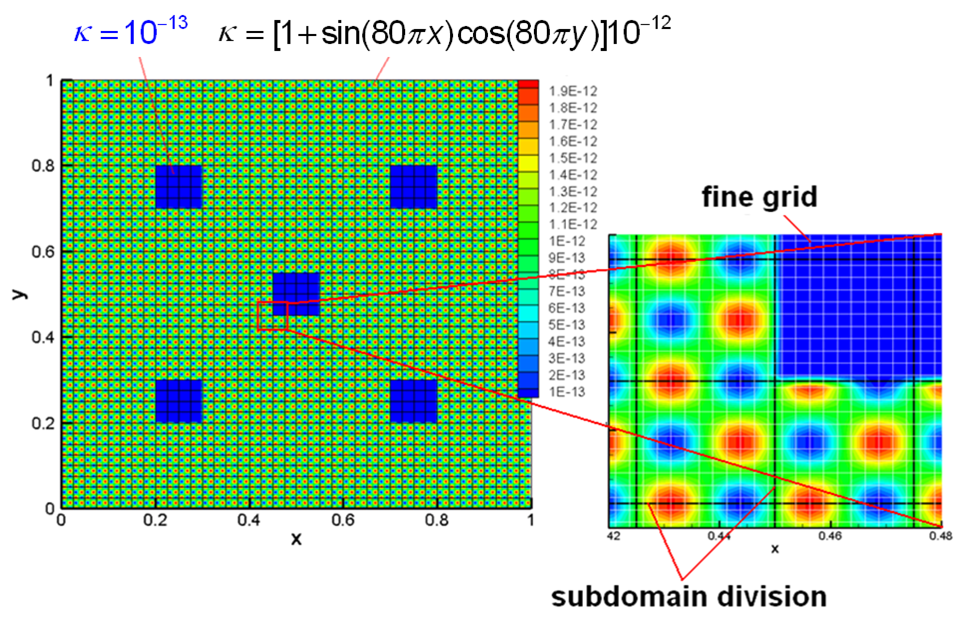}
  \caption{Distribution of the permeability $\kappa(\vec x)$, $\kappa_{\rm const}=10^{-13}$. }
  \label{fig:kappa}
\end{figure}

We choose a two-dimensional 1 m$\times$1 m domain with periodic boundary conditions and $\epsilon=0.8$, $\rho_0=1000$ kg m$^{-3}$, $\nu=2\times10^{-6}$ m$^2$ s$^{-1}$. We set $\nu_{\rm eff}=0$ m$^2$ s$^{-1}$ by using $\tau=0.5$ in the simulations of both fine and coarse grids and also in the calculation of $\mbox{\boldmath$\kappa$}^*(\vec x)$. In order to have obvious variations in the results of the coarse-grid simulation, a nonuniform external force $\vec G=(\sin\pi x, \sin\pi y)$ m s$^{-2}$ is used and the distribution of permeability $\kappa(\vec x)$ in Fig. \ref{fig:kappa} is set according to Eq. \eqref{eq:finekappa-fiveblock} such that the distribution of $\mbox{\boldmath$\kappa$}^*(\vec x)$ is nonuniform.

\begin{equation}\label{eq:finekappa-fiveblock}
    \begin{cases}\kappa=\kappa_{\rm const}, & 0.45\le x,y\le0.55 \\
    \kappa=\kappa_{\rm const}, & 0.2\le x,y\le0.3 \\
    \kappa=\kappa_{\rm const}, & 0.7\le x,y\le0.8 \\
    \kappa=\kappa_{\rm const}, & 0.2\le x\le0.3, 0.7\le y\le0.8 \\
    \kappa=\kappa_{\rm const}, & 0.7\le x\le0.8, 0.2\le y\le0.3 \\
    \kappa=10(1+\sin(80x\pi)\cos(80y\pi))\kappa_{\rm const}, & {\rm elsewhere},
    \end{cases}
\end{equation}
where $\kappa_{\rm const}=10^{-13}$ m$^2$. $\Delta x_{\rm fine}=0.0025$ m and $\Delta t_{\rm fine}=0.000025$ s in the fine-grid simulation. The number of fine points is 400$\times$400 inside the whole computational domain which is divided uniformly into 40$\times$40 subdomains. The averaged results over each set of $10\times10$ fine points located inside the same subdomain are computed and used to verify the results of the coarse-grid simulation.

We have $\kappa^*_{xx}=\kappa^*_{yy}=\kappa_{\rm const}$ and  $\kappa^*_{yx}=\kappa^*_{xy}=0$ inside the subdomains, where $\kappa\equiv\kappa_{\rm const}$. For subdomains with $\kappa=10(1+\sin(80x\pi)\cos(80y\pi))\kappa_{\rm const}$, we use $\vec G_{\rm const}=(2, 0)$ m s$^{-2}$ to drive flow and get $\kappa^*_{xx}=8.485\kappa_{\rm const}$ and $\kappa^*_{yx}=0$. The symmetric property of $\kappa(\vec x)$ inside the subdomain implies that $\kappa^*_{yy}=\kappa^*_{xx}$ and $\kappa^*_{xy}=\kappa^*_{yx}$. We define a scalar $\kappa^*$ as the average value over all diagonal components of $\mbox{\boldmath$\kappa$}^*$ and for all subdomains we have $\mbox{\boldmath$\kappa$}^*=\kappa^*\mathbf{I}$, where $\mathbf{I}$ is the identity tensor. Now, Eq. \eqref{eq:f_m depends on u and tensor k}, which is a general formula in the coarse-grid simulations, can be replaced by Eq. \eqref{eq:f_m depends on u}, where we change $\kappa$ to $\kappa^*$. In the case of $\mbox{\boldmath$\kappa$}^*\equiv\kappa^*\mathbf{I}$, the algorithm in the coarse-grid simulation is the same as in the fine-grid simulation (see Section \ref{ss:simple LBM algorithm}) but they use different scalar permeability distributions, namely $\kappa^*$ and $\kappa$, respectively. In the coarse-grid simulation, $\Delta x_{\rm coarse}=0.025$ m and $\Delta t_{\rm coarse}=0.00025$ s. The number of coarse points is $40\times40$ and the value of $\kappa^*$ assigned to each coarse point with index $(I, J)$ is:

\begin{equation}\label{eq:coarsekappa-fiveblock}
    \begin{cases}\kappa^*=\kappa_{\rm const}, & 19\le I,J\le22 \\
    \kappa^*=\kappa_{\rm const}, & 9\le I,J\le12 \\
    \kappa^*=\kappa_{\rm const}, & 29\le I,J\le32 \\
    \kappa^*=\kappa_{\rm const}, & 9\le I\le12, 29\le J\le32 \\
    \kappa^*=\kappa_{\rm const}, & 29\le I\le32, 9\le J\le12 \\
    \kappa^*=8.485\kappa_{\rm const}, & {\rm otherwise}.
    \end{cases}
\end{equation}

\begin{figure}
  \centering
  \includegraphics[width=0.32\textwidth]{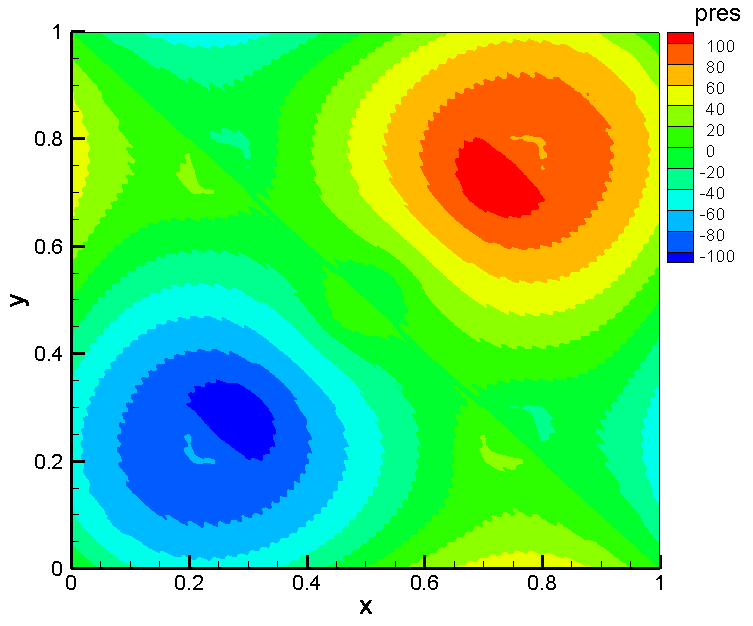}
  \includegraphics[width=0.32\textwidth]{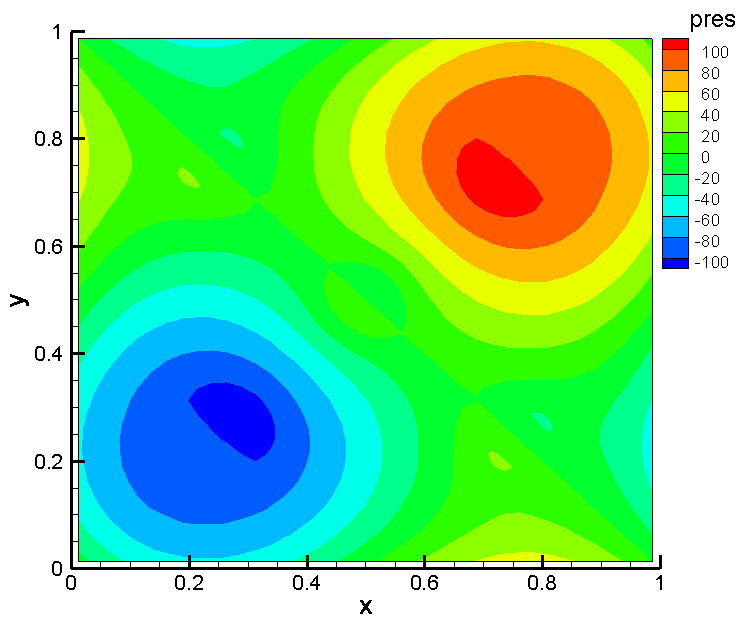}
  \includegraphics[width=0.32\textwidth]{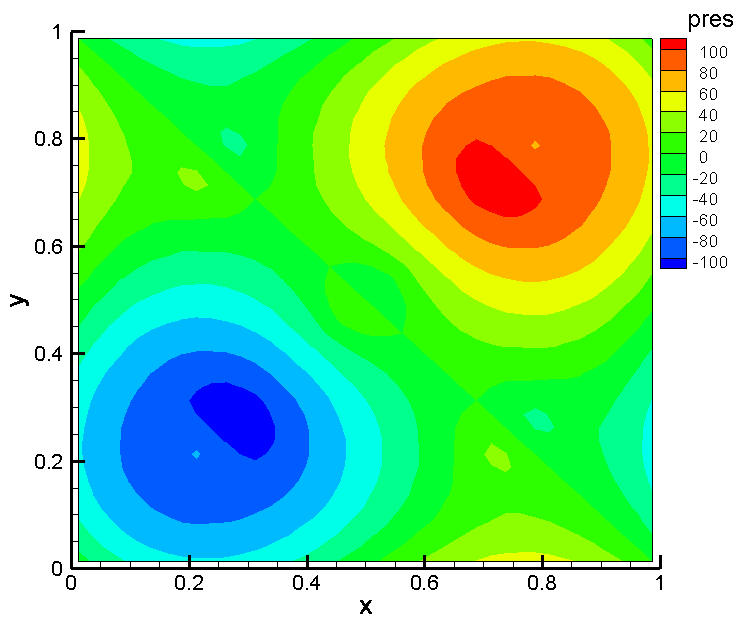} \\
  \includegraphics[width=0.32\textwidth]{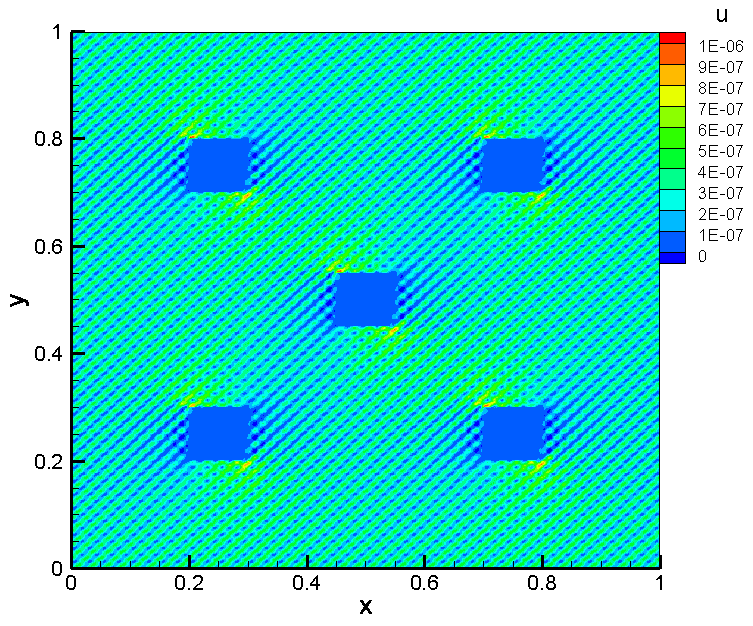}
  \includegraphics[width=0.32\textwidth]{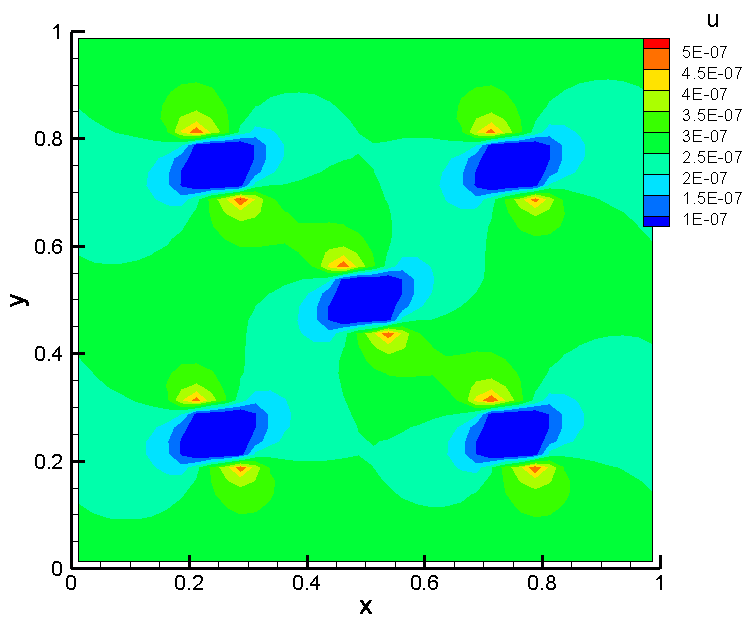}
  \includegraphics[width=0.32\textwidth]{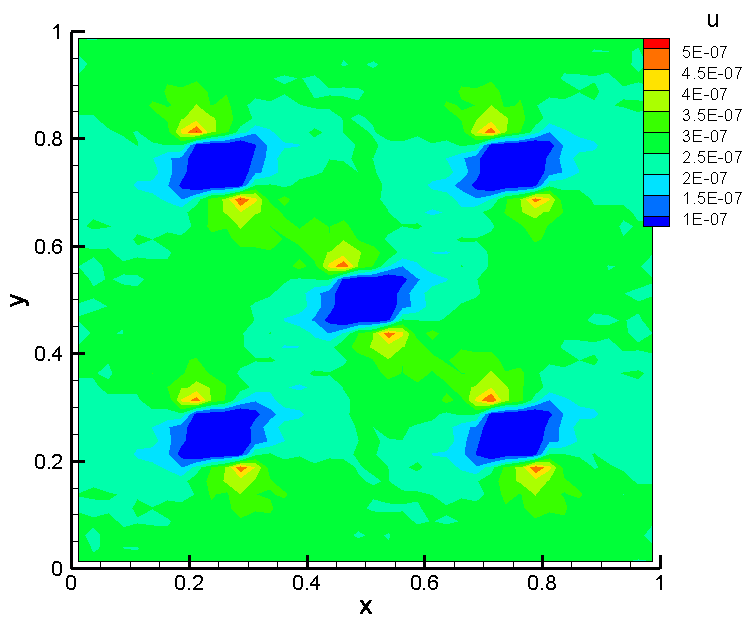} \\
  \includegraphics[width=0.32\textwidth]{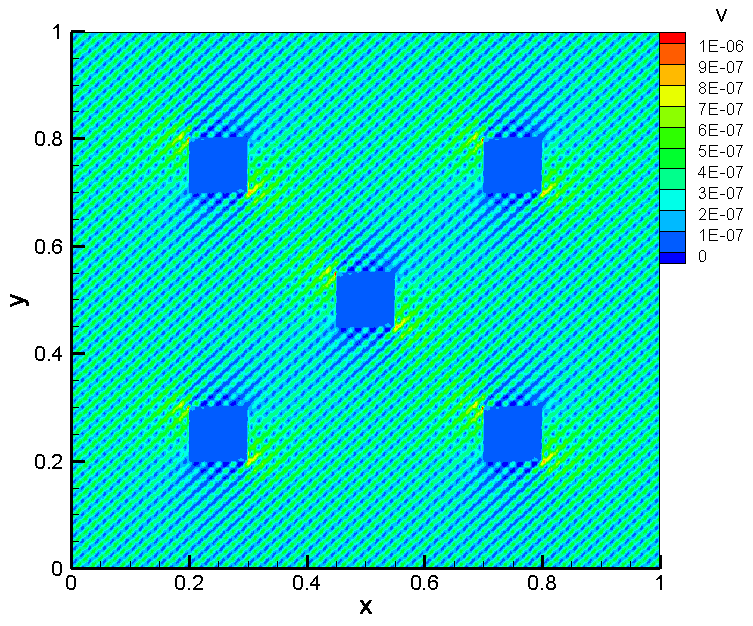}
  \includegraphics[width=0.32\textwidth]{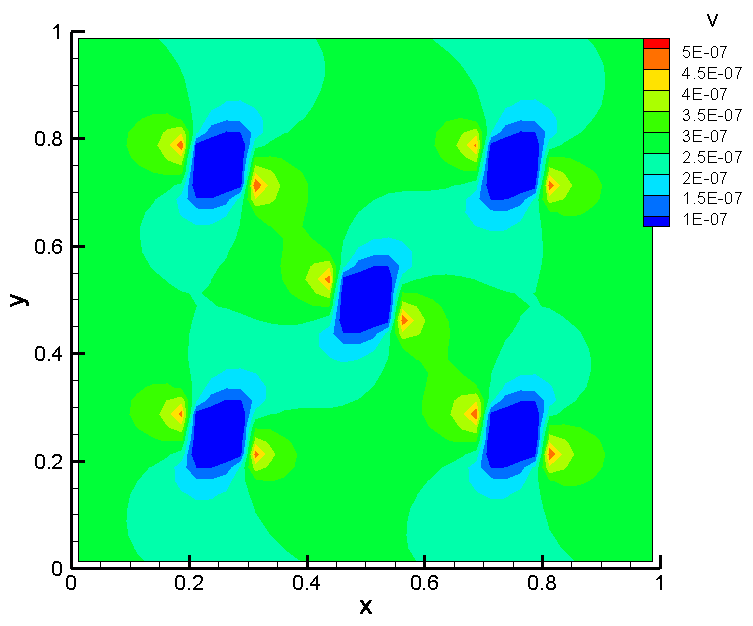}
  \includegraphics[width=0.32\textwidth]{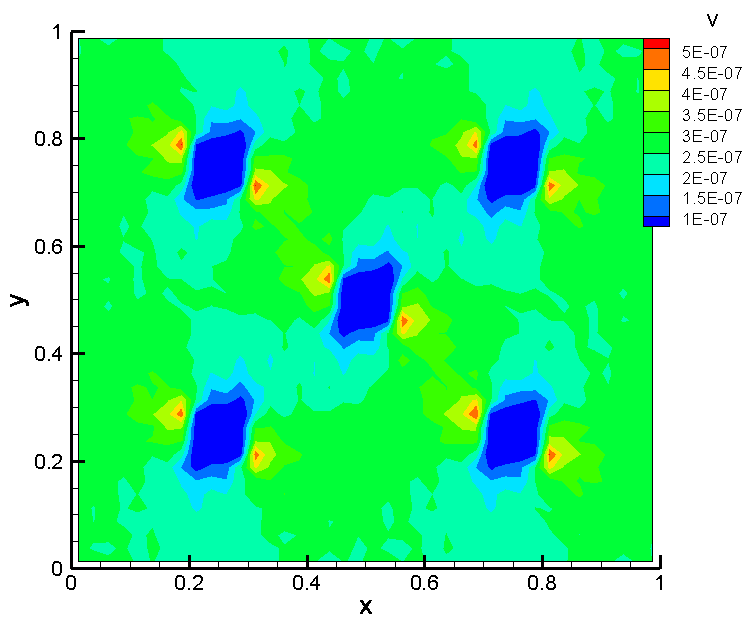}
  \caption{Comparisons of $p$, $u$ and $v$ between the fine-grid results (left), fine-grid averaged results (middle) and coarse-grid results using $\kappa^*$ (right), $\nu_{\rm eff}=0$ m$^2$ s$^{-1}$, $\vec G=(\sin\pi x, \sin\pi y)$ m s$^{-2}$, $\kappa_{\rm const}=10^{-13}$ m$^2$.}
  \label{fig:separate compare fiveblock}
\end{figure}

\begin{figure}
  \centering
  \includegraphics[width=0.32\textwidth]{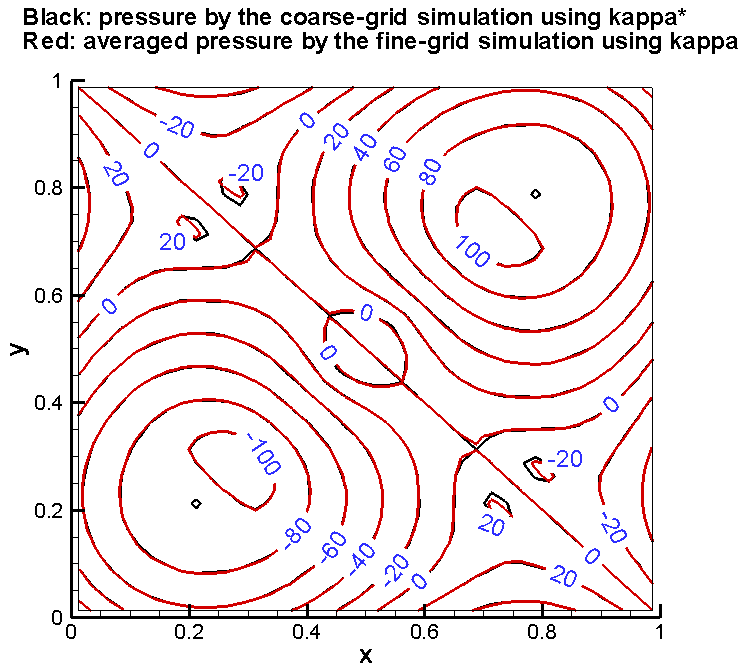}
  \includegraphics[width=0.32\textwidth]{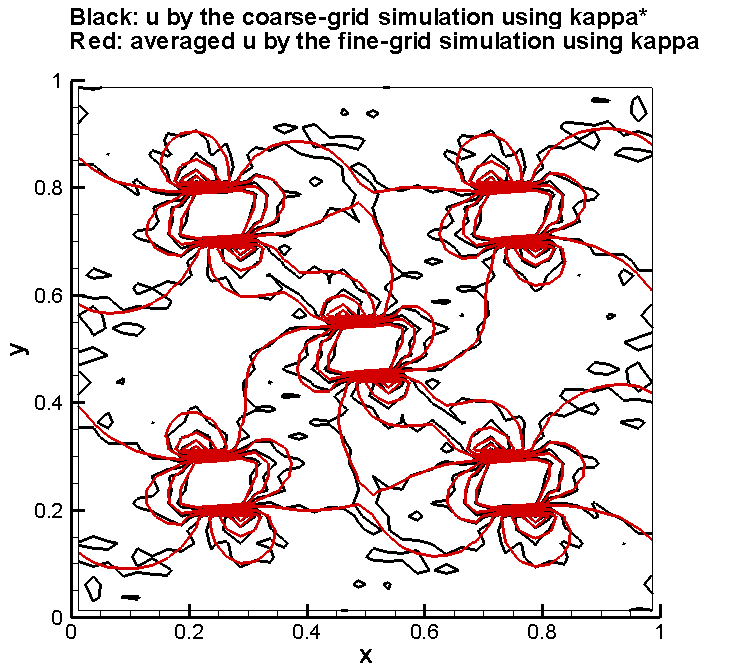}
  \includegraphics[width=0.32\textwidth]{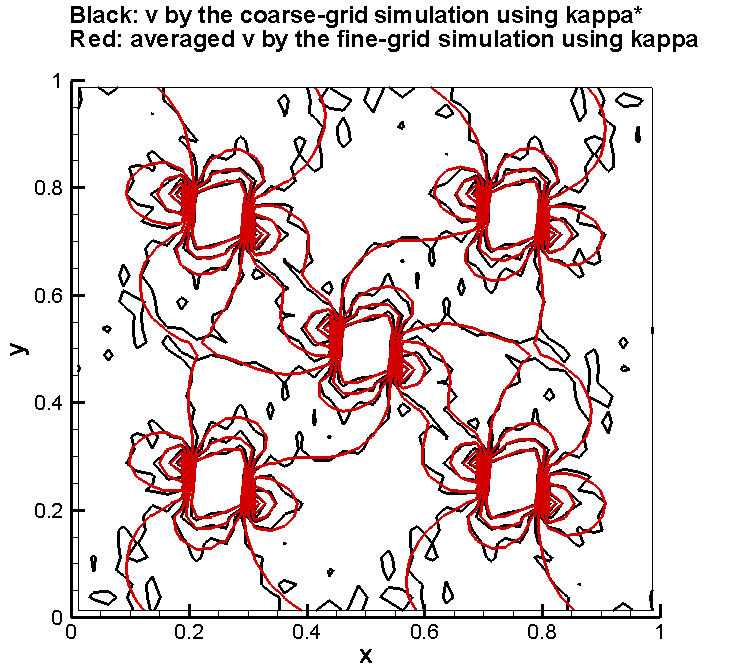}
  \caption{Detailed comparisons of $p$, $u$ and $v$ between the fine-grid averaged results and coarse-grid results using $\kappa^*$, $\nu_{\rm eff}=0$ m$^2$ s$^{-1}$, $\vec G=(\sin\pi x, \sin\pi y)$ m s$^{-2}$, $\kappa_{\rm const}=10^{-13}$ m$^2$.}
  \label{fig:overlap compare fiveblock}
\end{figure}

The distributions of the fine and coarse grids inside a representative area are given in Fig. \ref{fig:two grid model}. Figs. \ref{fig:separate compare fiveblock}-\ref{fig:overlap compare fiveblock} show that the agreement is very good between the two simulations using the fine and coarse grids, respectively.

\subsubsection{Simulations of Brinkman flows}\label{sss:Brinkman two grids}
\begin{figure}
  \centering
  \includegraphics[width=0.32\textwidth]{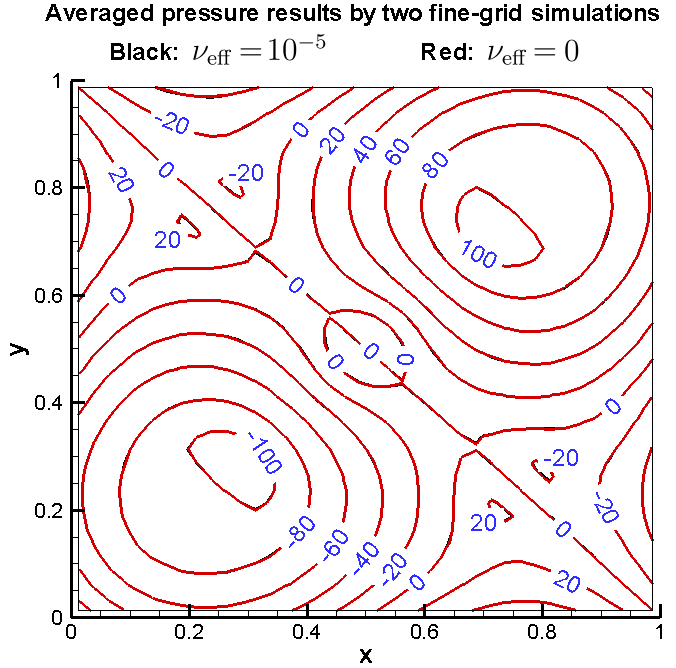}
  \includegraphics[width=0.32\textwidth]{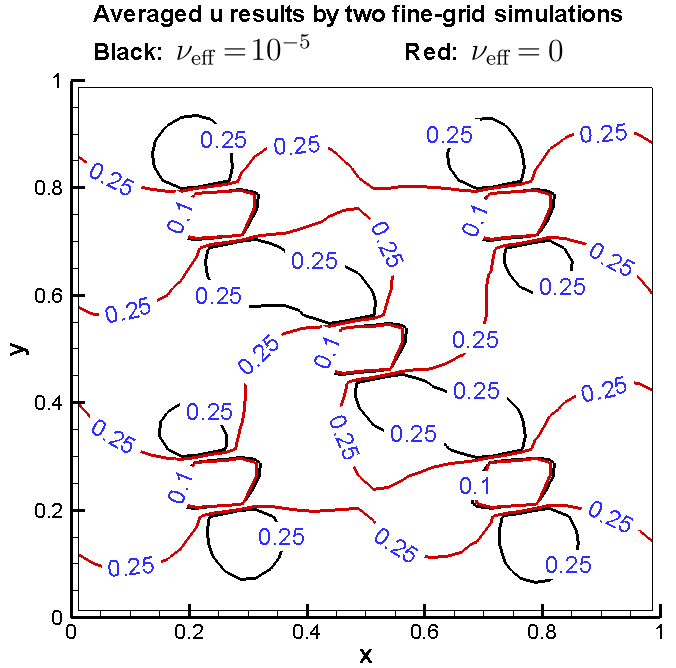}
  \includegraphics[width=0.32\textwidth]{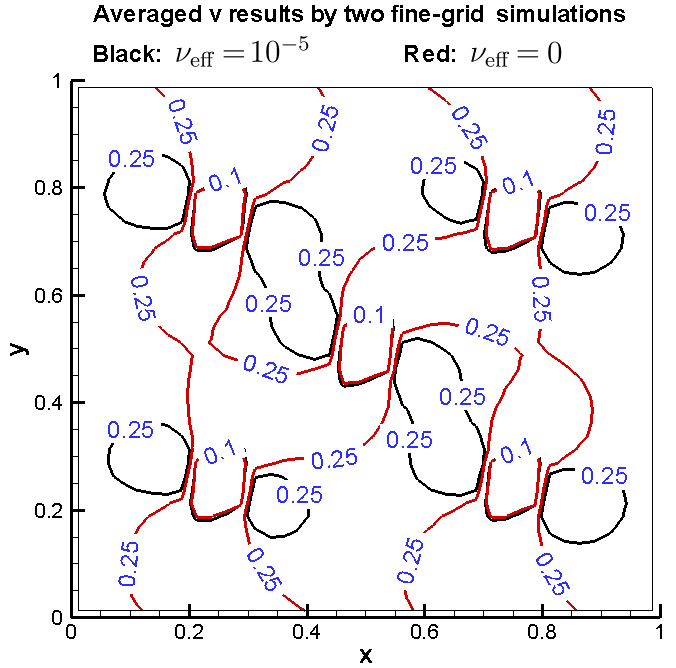}
  \caption{Comparisons of $p$, $u$ and $v$ between two fine-grid simulations using $\nu_{\rm eff}=10^{-5}$ and 0 m$^2$ s$^{-1}$, respectively, $\vec G=(\sin\pi x, \sin\pi y)$ m s$^{-2}$, $\kappa_{\rm const}=10^{-7}$ m$^2$.}
  \label{fig:difference due to nue}
\end{figure}

\begin{figure}
  \centering
  \includegraphics[width=0.32\textwidth]{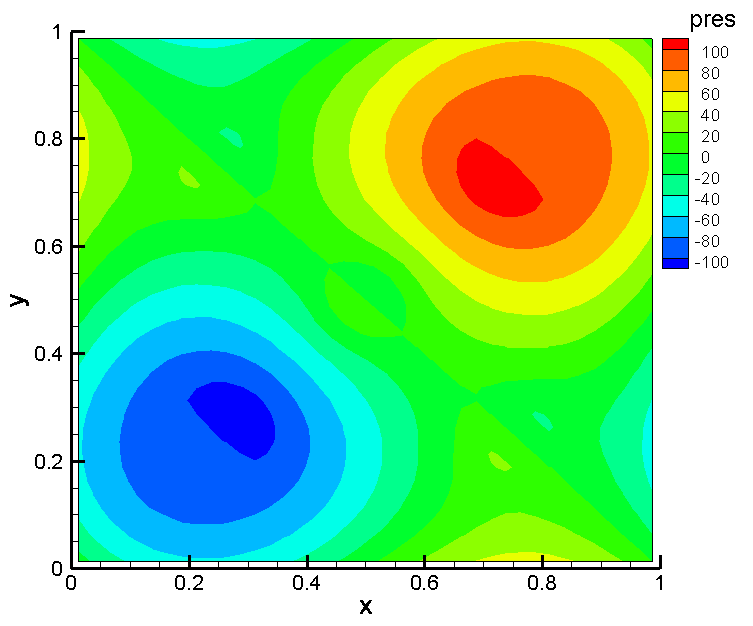}
  \includegraphics[width=0.32\textwidth]{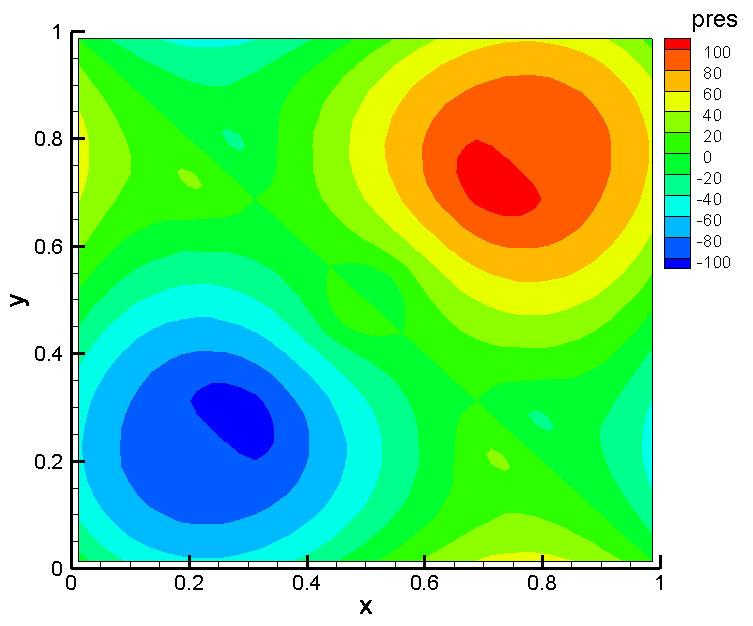}
  \includegraphics[width=0.32\textwidth]{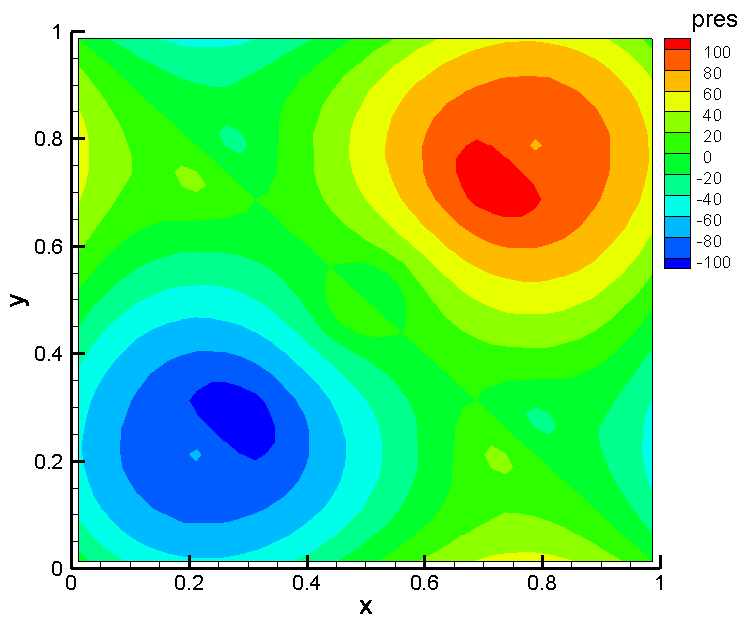} \\
  \includegraphics[width=0.32\textwidth]{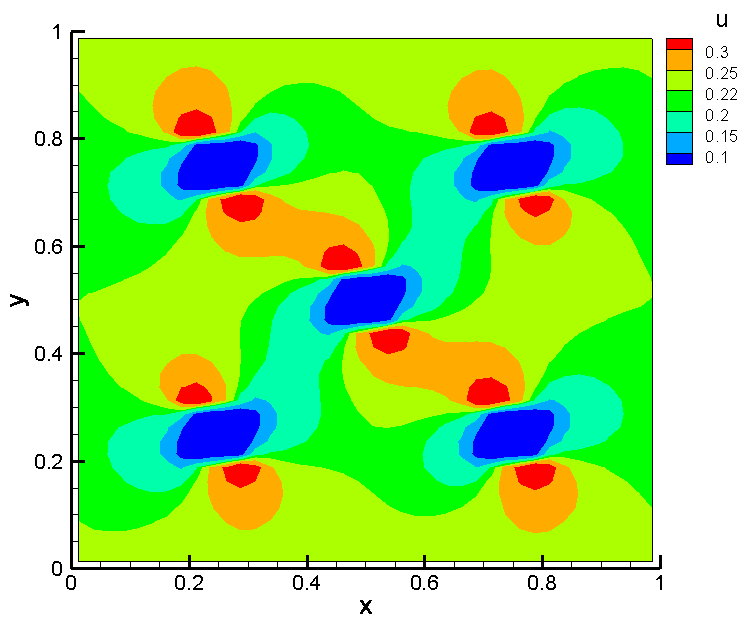}
  \includegraphics[width=0.32\textwidth]{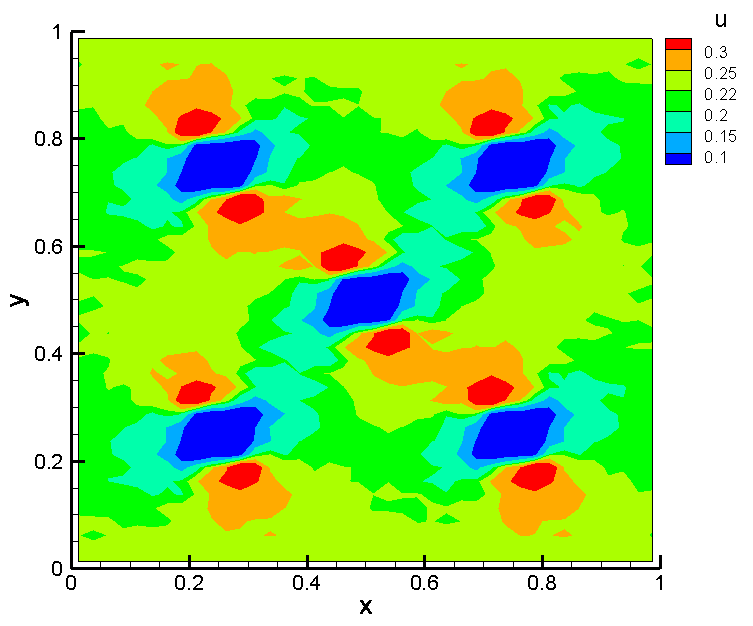}
  \includegraphics[width=0.32\textwidth]{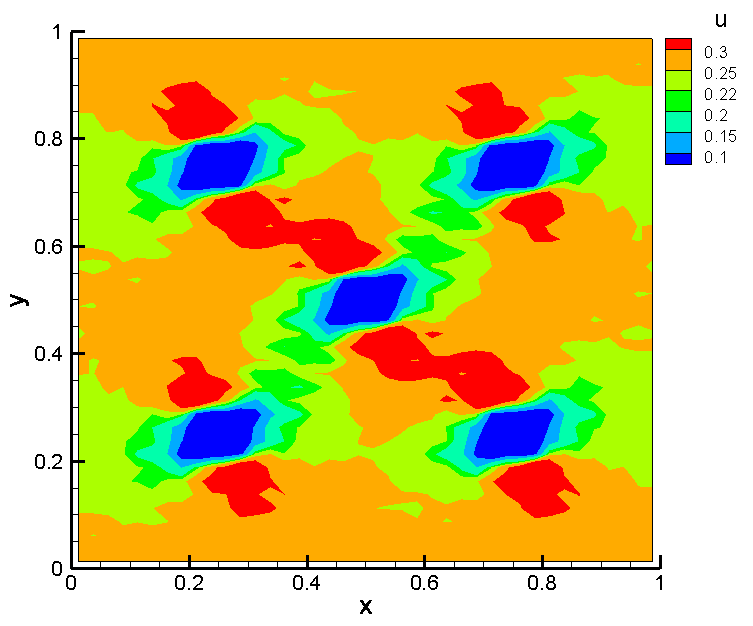} \\
  \includegraphics[width=0.32\textwidth]{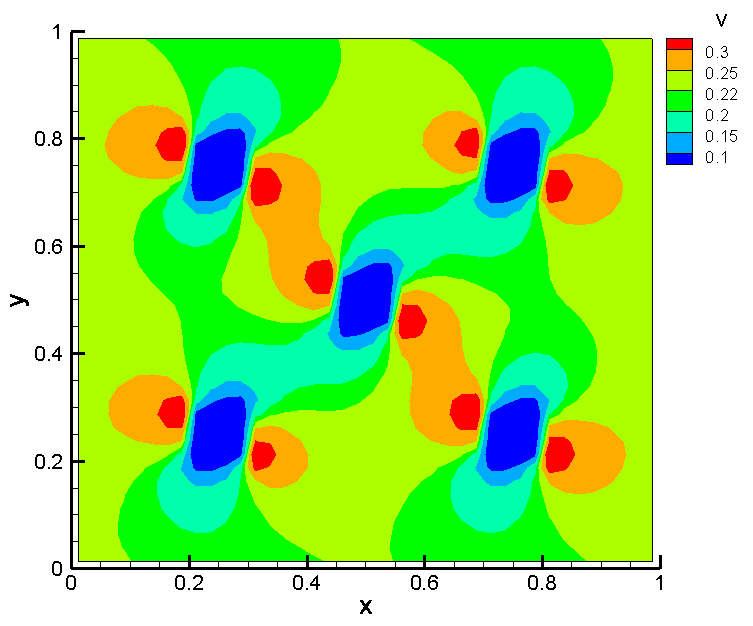}
  \includegraphics[width=0.32\textwidth]{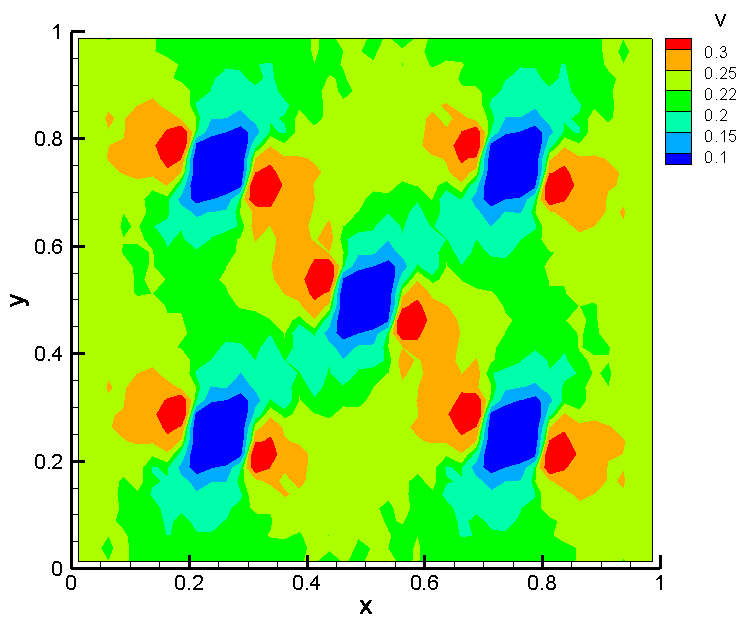}
  \includegraphics[width=0.32\textwidth]{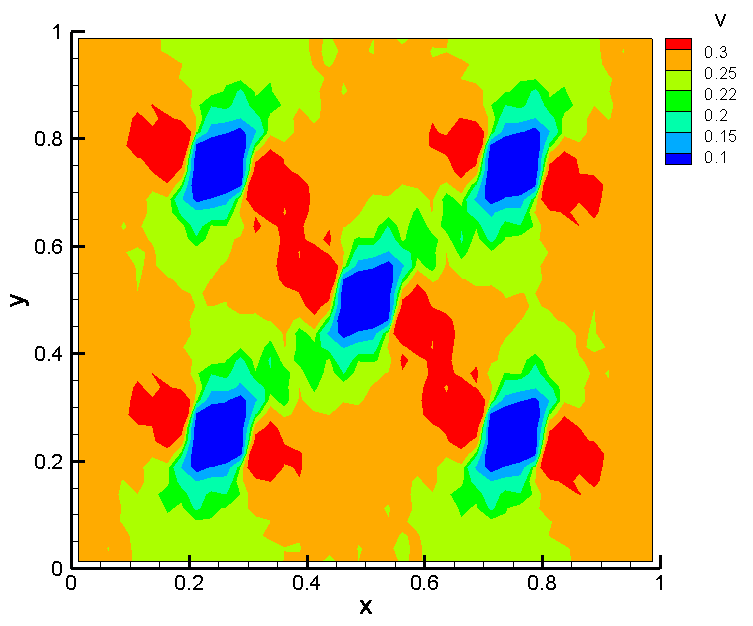}
  \caption{Comparisons of $p$, $u$ and $v$ between the fine-grid averaged results (left), coarse-grid results using $\kappa^*(\kappa, \nu_{\rm eff})$ (middle) and $\kappa^{*,\rm err}(\kappa)$ (right), $\nu_{\rm eff}=10^{-5}$ m$^2$ s$^{-1}$, $\vec G=(\sin\pi x, \sin\pi y)$ m s$^{-2}$, $\kappa_{\rm const}=10^{-7}$ m$^2$.}
  \label{fig:separate compare fiveblock kappa1}
\end{figure}

\begin{figure}
  \centering
  \includegraphics[width=0.32\textwidth]{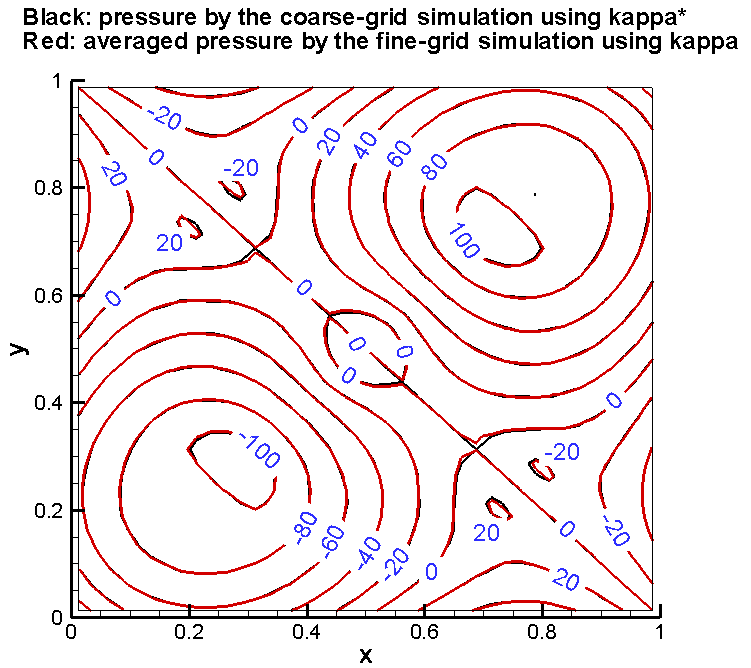}
  \includegraphics[width=0.32\textwidth]{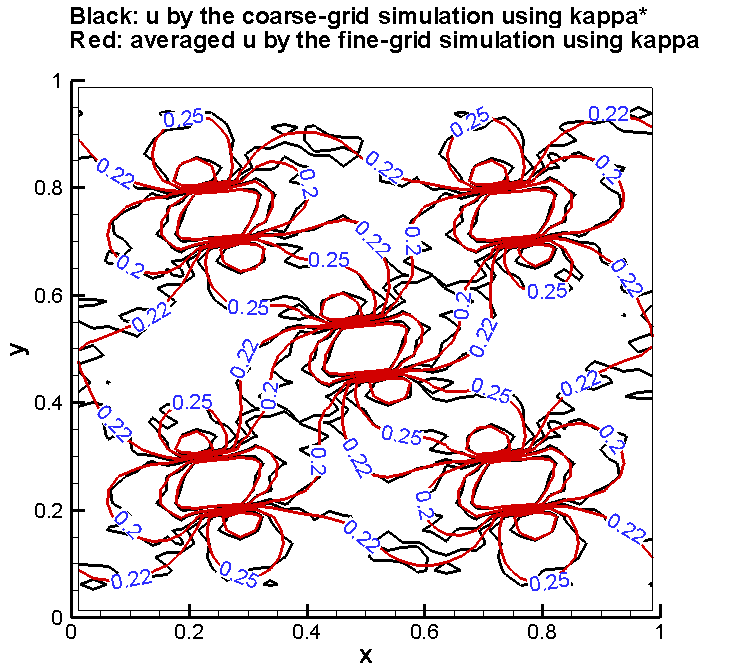}
  \includegraphics[width=0.32\textwidth]{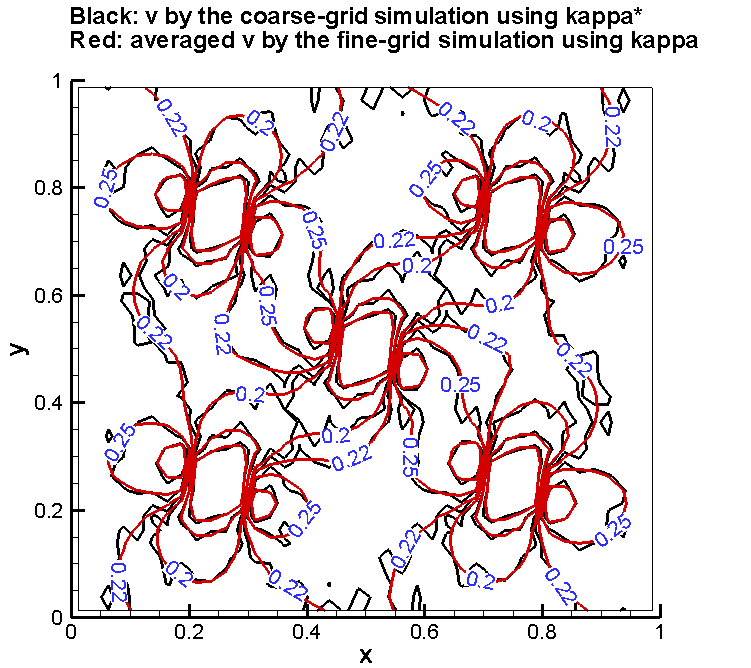}
  \caption{Detailed comparisons of $p$, $u$ and $v$ between the fine-grid averaged results and coarse-grid results using $\kappa^*(\kappa, \nu_{\rm eff})$, $\nu_{\rm eff}=10^{-5}$ m$^2$ s$^{-1}$, $\vec G=(\sin\pi x, \sin\pi y)$ m s$^{-2}$, $\kappa_{\rm const}=10^{-7}$ m$^2$.}
  \label{fig:overlap compare fiveblock kappa1}
\end{figure}

The physical problem studied here is similar to that described in Section \ref{sss:Darcy two grids}. The differences are that we increase the value of $\kappa_{\rm const}$ to be $\kappa_{\rm const}=10^{-7}$ m$^2$ (cf.  $\kappa_{\rm const}=10^{-13}$ m$^2$ in Section \ref{sss:Darcy two grids}) and set $\nu_{\rm eff}=10^{-5}$ m$^2$ s$^{-1}$ such that the contribution by the viscosity term $\nu_{\rm eff}\Delta\vec u$ is remarkable as shown in Fig. \ref{fig:difference due to nue}, which shows the comparison between the averaged results of two fine-grid simulations using $\nu_{\rm eff}=10^{-5}$ and 0 m$^2$ s$^{-1}$, respectively. In this regime, flow in some regions is close to Stokes flow while in other regions, flow is close to Darcy flow. Since $\nu_{\rm eff}$ is nonzero here, we let $\nu_{\rm eff}=10^{-5}$ m$^2$ s$^{-1}$ when computing $\mbox{\boldmath$\kappa$}^*$ and get $\kappa^*_{xx}(\kappa, \nu_{\rm eff})=\kappa^*_{yy}(\kappa, \nu_{\rm eff})=7.367\kappa_{\rm const}$ and $\kappa^*_{yx}(\kappa, \nu_{\rm eff})=\kappa^*_{xy}(\kappa, \nu_{\rm eff})=0$ for subdomains with $\kappa=10(1+\sin(80x\pi)\cos(80y\pi))\kappa_{\rm const}$. For subdomains with $\kappa=\kappa_{\rm const}$, $\kappa^*_{xx}(\kappa, \nu_{\rm eff})=\kappa^*_{yy}(\kappa, \nu_{\rm eff})=\kappa_{\rm const}$ and $\kappa^*_{yx}(\kappa, \nu_{\rm eff})=\kappa^*_{xy}(\kappa, \nu_{\rm eff})=0$. Thus, we have $\mbox{\boldmath$\kappa$}^*(\kappa, \nu_{\rm eff})$ equal to $7.367\kappa_{\rm const}\mathbf{I}$ or $\kappa_{\rm const}\mathbf{I}$. As discussed in Section \ref{sss:Darcy two grids}, we can use a scalar distribution of $\kappa^*(\kappa, \nu_{\rm eff})$, which is equal to $7.367\kappa_{\rm const}$ or $\kappa_{\rm const}$, in the coarse-grid simulation. We use $\Delta x_{\rm fine}=0.0025$ m, $\Delta t_{\rm fine}=0.000025$ s and $\tau_{\rm fine}=0.50012$ in the fine-grid simulation, and use $\Delta x_{\rm coarse}=0.025$ m, $\Delta t_{\rm coarse}=0.00025$ s and $\tau_{\rm coarse}=0.500012$ in the coarse-grid simulation. Figs. \ref{fig:separate compare fiveblock kappa1}-\ref{fig:overlap compare fiveblock kappa1} show that the agreement of the coarse-grid simulation using $\kappa^*(\kappa, \nu_{\rm eff})$ with the fine-grid simulation is very good. In addition, we set $\nu_{\rm eff}=0$ m$^2$ s$^{-1}$ when computing $\mbox{\boldmath$\kappa$}^{*,\rm err}$ and get $\kappa_{xx}^{*,\rm err}(\kappa)=\kappa_{yy}^{*,\rm err}(\kappa)=8.485\kappa_{\rm const}$ and $\kappa_{yx}^{*,\rm err}(\kappa)=\kappa_{xy}^{*,\rm err}(\kappa)=0$ for subdomains with $\kappa=10(1+\sin(80x\pi)\cos(80y\pi))\kappa_{\rm const}$. For subdomains with $\kappa=\kappa_{\rm const}$, $\kappa^{*,\rm err}_{xx}(\kappa)=\kappa^{*,\rm err}_{yy}(\kappa)=\kappa_{\rm const}$ and $\kappa^{*,\rm err}_{yx}(\kappa)=\kappa^{*,\rm err}_{xy}(\kappa)=0$, where we still use the superscript 'err' since the local simulation procedure with $\nu_{\rm eff}=0$ m$^2$ s$^{-1}$ is wrong although the obtained $\mbox{\boldmath$\kappa$}^{*,\rm err}(\kappa)$ is the same as the above $\mbox{\boldmath$\kappa$}^*(\kappa, \nu_{\rm eff})$. Now, we have $\mbox{\boldmath$\kappa$}^{*,\rm err}(\kappa)$ equal to $8.485\kappa_{\rm const}\mathbf{I}$ or $\kappa_{\rm const}\mathbf{I}$. We use a scalar distribution of $\kappa^{*,\rm err}(\kappa)$, which is equal to $8.485\kappa_{\rm const}$ or $\kappa_{\rm const}$, in another coarse-grid simulation. Note that the difference between $\kappa^{*,\rm err}(\kappa)$ and $\kappa^*(\kappa, \nu_{\rm eff})$ is distinct in subdomains with nonuniform $\kappa(\vec x)$. The results of the coarse-grid simulation using $\kappa^{*,\rm err}(\kappa)$ are also given in Fig. \ref{fig:separate compare fiveblock kappa1} which shows that the deviation of the coarse-grid simulation using $\kappa^{*,\rm err}(\kappa)$ from the fine-grid simulation is remarkable. Thus, the previous computation of $\mbox{\boldmath$\kappa$}^*(\kappa, \nu_{\rm eff})$ as the effective permeability using nonzero $\nu_{\rm eff}$ is accurate for upscaling the Brinkman equation.

%
\section{Conclusions}

Pore-scale flows are routinely modeled by the LBM simulations due to their ability to handle complex geometries and physics. However, LBM simulations become very expensive as one uses large REVs. In this paper, we propose a upscaled LBM algorithm to model flows at coarse
scales with a reduced computational complexity. The effective properties are computed by a local upscaling scheme. In this scheme, the local fine-grid simulations are performed and their results are averaged over the local region to compute effective properties. Effective properties are used in a coarse-grid LBM algorithm to perform the simulations at larger scales. The coarse-grid LBM simulation using the computed effective permeability agrees very well with the fine-grid LBM simulation. In addition, simulation results show that the coarse-grid LBM simulation will deviate significantly from the fine-grid LBM simulation if the effective permeability is computed by neglecting the viscosity term in modeling Brinkman flows.

Although the results presented in this paper are encouraging, there
is scope for further exploration of some of the underlying
approaches. As our intent here was to demonstrate that coarse
scale information could be effectively used to design upscaled LBM
representations, we did not consider challenging
heterogeneous cases with high-contrast permeability. It is known
(e.g., \cite{Efendiev2009:MsFEM} and \cite{Qin2010:SPE}) that the presence of high heterogeneities,
such as channels and high contrast, will cause a decrease in the accuracy
of upscaling methods for Darcy flow problems. Similarly, we expect that
our upscaled LBM algorithm will require an additional treatment to
handle highly heterogeneous cases. These treatments can include
oversampling, local-global, or global techniques or possibly upscaled
techniques. Some of these treatments can be easily incorporated
into our new upscaled LBM framework.

\section{Acknowledgements}

We would like to acknowledge Victor Calo for his helpful discussion on the physical implications of the models. Also, we would like to thank Oleg Iliev for his helpful insights on the upscaling of porous media and validation of the computational results.

\end{document}